\documentclass[final,5p,times,twocolumn,authoryear]{elsarticle}

\usepackage{amssymb}
\usepackage{amsmath}

\usepackage{aas-macros}
\usepackage{booktabs}
\usepackage{xcolor}
\usepackage{hyperref}

\usepackage{newtxtext,newtxmath}
\usepackage{supertabular,lscape,epsfig}
\usepackage{graphicx}%
\usepackage{multirow}%
\usepackage{amsmath,amsfonts}%
\usepackage{mathrsfs}%
\usepackage[title]{appendix}%
\usepackage{xcolor}%
\usepackage[table]{xcolor}
\usepackage{textcomp}%
\usepackage{manyfoot}%
\usepackage{booktabs}%
\usepackage{algorithm}%
\usepackage{algorithmicx}%
\usepackage{algpseudocode}%
\usepackage{listings}%
\usepackage{hyperref}
\usepackage[nopatch]{microtype}
\usepackage{tikz}
\usepackage{tabularx}
\usepackage{longtable}
\usepackage{array}
\usepackage{mathrsfs}
\usepackage{fix-cm}
\usepackage[export]{adjustbox}
\usepackage{rotating}
\usepackage{collcell}   
\usepackage{inputenc}

\usepackage[T1]{fontenc}

\DeclareRobustCommand{\VAN}[3]{#2}
\let\VANthebibliography\thebibliography
\def\thebibliography{\DeclareRobustCommand{\VAN}[3]{##3}\VANthebibliography}

\usepackage{graphicx}	
\usepackage{amsmath}	

\hypersetup{
    colorlinks=false,    
    linkcolor=red,      
    citecolor=blue,     
    urlcolor=cyan,      
    filecolor=magenta   
}

\journal{Advances in Space Research}

\newcommand\arcsec{\hbox{$^{\prime\prime}$}}

\begin{document}
\begin{frontmatter}

\title{Photoionization Modeling of Planetary Nebulae in the Galactic Bulge}



\author[1,2]{Nazım Aksaker} \ead{naksaker@cu.edu.tr} \cortext[cor1]{Corresponding author}
\author[2]{Aleyna Demirci}
\author[2]{Nurullah Erzincan}
\author[1,3]{Aysun Akyuz}

\affiliation[1]{%
    organization={Çukurova University},
    addressline={Space Science and Solar Energy Research and Application Center (UZAYMER)},
    city={Adana},
    postcode={01330},
    country={Türkiye}}
\affiliation[2]{%
    organization={Çukurova University},
    addressline={Adana Organised Industrial Zones Vocational School of Technical Science},
    city={Adana},
    postcode={01410},
    country={Türkiye}}

\affiliation[3]{%
    organization={Çukurova University},
    addressline={Department of Physics},
    city={Adana},
    postcode={01330},
    country={Türkiye}}

\begin{abstract}
In this study, we present the results of photoionization modeling for 124 planetary nebulae (PNe) in the Galactic bulge. Utilizing the {\scshape cloudy} code, we derived the effective temperatures (T$_{eff}$) of the central stars, with a peak distribution around $\sim$ 100,000 K, and luminosities clustering around $\sim$ 3,000 L$\odot$. The inner radii of the ionized regions range from 0.003 to 0.31 pc, with nebula diameters varying from 1.8" to 34", averaging 7\% larger than the observed visible diameters. 
Elemental abundances for Helium, Nitrogen, Oxygen, Neon, Sulphur, Chlorine and Argon relative to hydrogen derived from the models show consistency within 0.5 dex, with notable variations in Sulphur, Nitrogen, and Chlorine. The study also compares elemental abundances from photoionization models with previous observations, showing overall good agreement, particularly for Cl/H, but notable discrepancies in He/H and S/H ratios. The models' goodness of fit, quantified by $\chi^{2}$ values, varies widely, with higher values linked to discrepancies in WISE photometric data. The evolutionary tracks of the central stars from H-R diagrams suggest progenitor masses ranging from 0.8 to 4.2 $M_\odot$ and progenitor final masses between 0.53 and 0.87 $M_\odot$, indicating significant mass loss during evolution. These PNe have post-AGB ages ranging from 150 to 20,000 years, consistent with the Galactic PNe distribution. Most of the PNe are in an intermediate evolutionary stage with larger nebular sizes. Our results provide the most comprehensive photoionization modeling to date, with key implications for central stars of PNe, gas, and dust.
\end{abstract}

\end{frontmatter}


\section{Introduction}

Intermediate-mass stars (0.8–8 M$\odot$), evolve into the Asymptotic Giant Branch (AGB) phase at the end of their life cycle. During this phase, the star has an inert carbon-oxygen core and two active nuclear burning shells, one fusing helium into carbon and the other converting hydrogen into helium. As the star sheds its outer layers through strong stellar winds, its core contracts and temporarily reignites, ionizing the expelled gas. This ionized gas emits light, forming a glowing shell known as a planetary nebula (PN) \citep{2024Galax..12...39K}. During the formation of PNe, post-AGB stars expel their outer layers over a period of up to 30,000 years, dispersing enriched material into the interstellar medium as they evolve into white dwarfs (WDs). The exposed core, now the central star of the PN (CSPN), is typically a hot WD with an effective temperature between 60,000 and 200,000 K \citep{2004A&A...426..779M}. Despite having a diameter roughly half that of the Sun, these CSPNe exhibit high luminosities, ranging from 100 to 10$^5$ L${\odot}$ \citep{2008ApJ...674..954B, 2024A&A...688A..36T}. 

As key tracers of the Galactic structure and chemical evolution, PNe provide crucial insights into the process of interstellar enrichment \citep{2022PASP..134b2001K} (KH22). Elemental abundances of Helium (He), Nitrogen (N), Oxygen (O), Neon (Ne), Sulphur (S), Chlorine (Cl) and Argon (Ar) vary among the Galactic components (bulge, disk and halo) and include numerous PNe that are particularly suitable for detailed analysis (\cite{2006MNRAS.373...79P}; KH22). Previous studies have often overlooked the large population of PN in Galactic components, focusing instead on limited samples. 
Recent findings indicate lower abundances that deviate from the radial gradient of the disk, supporting updated chemical evolution models, with no evidence of extreme metallicities \citep{2015A&A...583A..71P, 2024MNRAS.527.6363T} (Tan24).
Spectroscopic analyses of PNe in the Galactic bulge reveal an older, light-metal-rich population with a lower N/O ratio, which is indicative of the metallicity of the progenitor stars compared to their counterparts in the disk \citep{2000A&A...353..543C, 2004MNRAS.349.1291E, 2009A&A...500.1089G, 2012IAUS..283..326C, 2015A&A...583A..71P}.

The properties of CSPN, such as mass and age, are crucial to understanding the characteristics of PNe.
Mass estimates for nearly 100 Galactic bulge PNe indicate a mean CSPN mass of 0.593$\pm$0.025 M$\odot$, lower than their Galactic disk counterparts \citep{1991A&A...246..221T}. A spectroscopic survey identified 492 confirmed or probable CSPNe, revealing that at least 30\% exhibit hydrogen-poor atmospheres and providing insight into their origin \citep{2011A&A...526A...6W}. The most comprehensive study in this field is \cite{2020A&A...640A..10W}, which provides a catalog of 620 CSPNe, including spectral classifications and key parameters such as luminosity, effective temperature, and surface gravity. Furthermore, \cite{2021MNRAS.506.5223J} investigated the CSPN binarity in 30 bulge PNe, providing periods and preliminary classifications (eclipsing, double degenerate, or irradiated systems) for likely binaries based on light-curve shapes. 

\cite{2023MNRAS.519.1049T} investigated the morphologies and CSPN in the Galactic bulge using deep FORS2 spectra and images of the 8.2 m VLT. Building on this database, Tan24 reported precise abundances for 124 Galactic bulge PNe, expanding the existing sample. The study refines interstellar reddening estimates, electron densities, temperatures, and chemical compositions, yielding a robust dataset. Their follow-up study \citep{2024ApJ...961L..47T} investigates the sulfur anomaly, attributed discrepancies to ionization correction factors rather than true deficiencies for the same PNe population. 

In this study, we investigated the central star, gas, and dust properties of 124 Galactic bulge PNe using the most comprehensive photoionization modeling to date. These PNe were selected from the Tan24 study, which did not explore these specific features.
This paper is organized as follows: Section \ref{sec:database} details the database, Section \ref{sec:cloudy} outlines the photoionization modeling process, Section \ref{sec:res} presents the results and discussion, and Section \ref{sec:conc} summarizes the conclusions.

\section{Database}
\label{sec:database}

The database comprises 124 PNe located within a $10^{\circ} \times 10^{\circ}$ region of the Galactic bulge, where the source positions are within $10^{\circ}$ of the Galactic center in both Galactic longitude and latitude \citep{2013MNRAS.435..975R}. Table \ref{T:obslog} presents the PNe along with their common names, coordinates, major diameters, R magnitudes (R$_{mag}$), and distances. The values in the first three columns are taken from the Hong Kong/Australian Astronomical Observatory/Strasbourg Observatory H-alpha Planetary Nebulae \footnote{http://hashpn.space}(HASH) database \citep{2016JPhCS.728c2008P}. The R$_{mag}$ comes from the catalog {\it USNO B1.0} \citep{2003AJ....125..984M}. We chose $R_{\mathrm{mag}}$ because it is available for all 124 PNe studied and enables direct comparison between them. The distances of PNe are obtained from the Gaia EDR3 database\footnote{https://vizier.cds.unistra.fr/viz-bin/VizieR-3?-source=I/352/gedr3dis}, while the distances for their central stars (CSPN) are adopted from \cite{2021A&A...656A.110C}.
Wide-field Infrared Survey Explorer (WISE) images were used for visualizing and modeling the PNe. The RGB images (R: W4 (22 µm), G: W3 (12 µm), B: W2 (4.6 µm)) of selected four PNe are presented in Fig. \ref{fig:wise}.

We utilized flux data for 124 PNe from Tan24, available as a supplementary zip file on the journal homepage, containing a CSV with 32 key emission lines (see Table \ref{flux_table}) and errors (3726–8046 {\AA}). Emission line fluxes are normalized to H$_{\beta}$ $\lambda$4861 {\AA}, which is set to 100. Here, the flux values have been corrected for reddening using the method of \cite{1999PASP..111...63F} with {\scshape pyneb}. Additionally, we utilized Table 7 of Tan24, which presents the total elemental abundances of He, N, O, Ne, S, Ar, and Cl (relative to H) for each PN, along with their associated uncertainties for the photoionization modeling. We also used Infrared Astronomical Satellite (IRAS, preferably), WISE, and radio flux values given in Table \ref{T:cloudy_in}. 

\section{Photoionization Modeling}
\label{sec:cloudy}

\begin{figure*}
\centering
\begin{tabular}{@{}c@{}c@{}}
\includegraphics[angle=0,width=1.0\columnwidth]{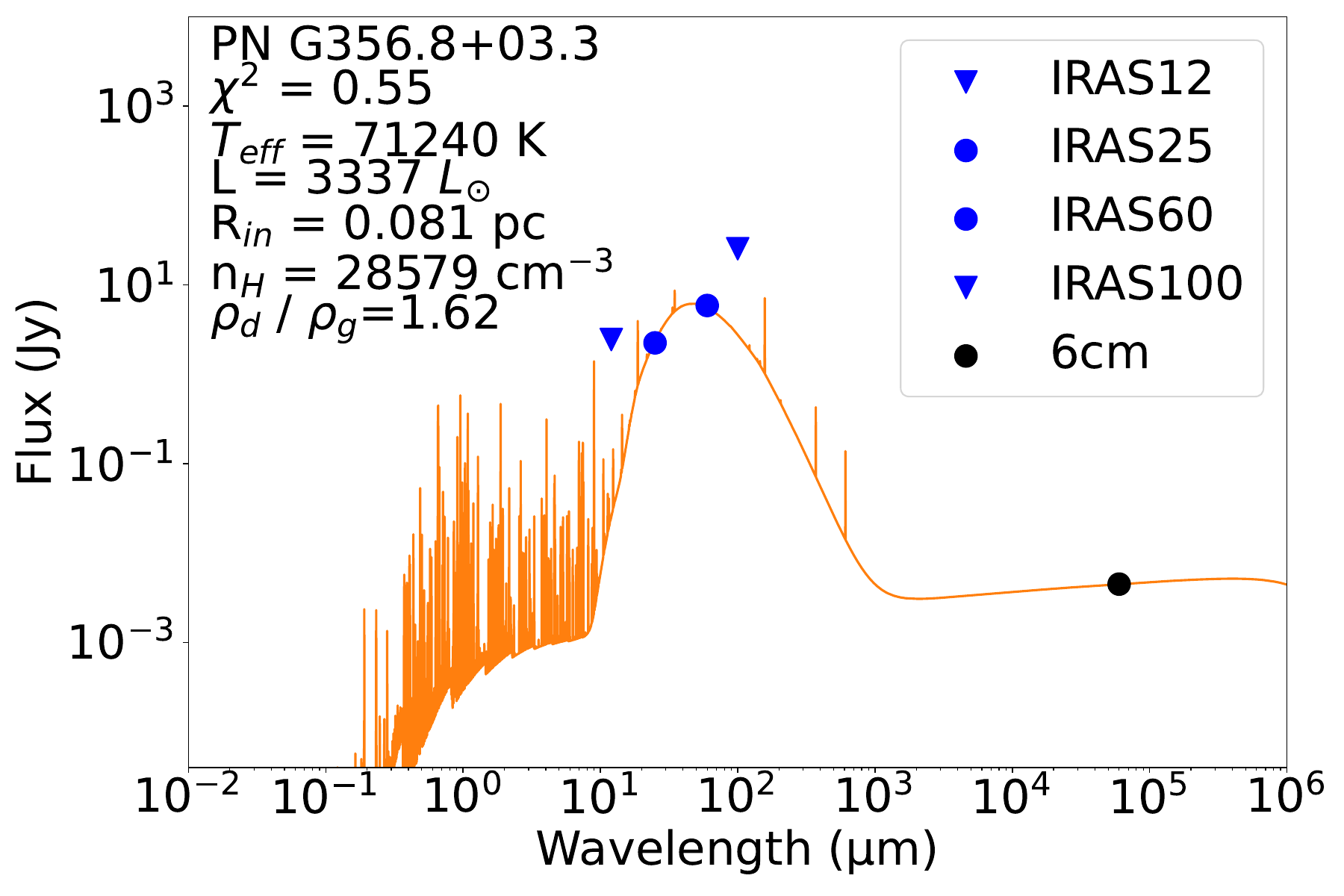}
\end{tabular}
\caption{SED of PN G356.8+03.3. The SED, generated from the best {\scshape cloudy} model with a $\chi^2 = 0.55$ among all PNe, is represented by the orange lines. Key parameters derived from the model are shown in the upper left corner. Blue circles denote IRAS data, the black circle represents radio data, and the blue inverted triangles indicate the upper limit of the IRAS data.}
\label{F:model}
\end{figure*}

Photoionization models of PNe are essential for determining key physical parameters such as central star temperature, luminosity, and gas and dust properties. This study introduces photoionization models for 124 PNe, contributing to the previous studies that analyze multiple PNe \citep{1999MNRAS.308..623V,2008ApJ...674..954B, 2024AcA....74..201A}. The photoionization modeling was conducted using {\scshape cloudy} (v.23.01; \citealp{2023RMxAA..59..327C}) and pycloudy \citep{2013ascl.soft04020M}, a Python package for managing photoionization models. pycloudy was utilized to model Spectral Energy Distributions (SEDs) and visualize structures. Fig. \ref{F:model} illustrates as an example for the best-fit {\scshape cloudy} model of PN G356.8+03.3 among the 124 PNe.

We followed the modeling techniques of \cite{1999MNRAS.308..623V} and \cite{ 2024AcA....74..201A} to analyze the selected PNe. The PN model assumes a homogeneous spherical shell of gas surrounding a hot central star, typically a white dwarf, emitting as a blackbody. Within the Strömgren radius, the gas density remains constant, while beyond this radius, it follows a 1/r$^{2}$ decline. 

In our {\scshape cloudy} models, graphite and silicate grains are assumed to be co-spatial and mixed, with their relative abundances determining their contribution to the overall dust emission. While many PNe are dominated by either carbon- or oxygen-rich dust, some PNe show dual-dust chemistry, occasionally with spatial segregation of the different dust types \citep{2009A&A...495L...5P, 2018MNRAS.473.4476G}. The mixed-grain assumption simplifies the modeling of the nebula’s infrared emission. While this may introduce some uncertainties in the derived dust temperatures and emissivities, it provides a reasonable estimate of the nebula’s integrated infrared properties.

Below, we present the details of the parameters used in the {\scshape cloudy} models. All parameters are constrained within lower and upper bounds, and commonly accepted values for PNe, as established in the literature, are adopted to reduce degeneracy.

The distance is a crucial parameter in {\scshape cloudy} modeling because it directly affects the luminosity and effective temperature (T$_{eff}$). To determine distances, we utilized data from the Gaia mission \citep{2021AJ....161..147B}. This mission provides a 3D map of the Milky Way and continuously improves distance estimates for PNe. A recent study by \cite{2021A&A...656A.110C} utilized Gaia distances to more accurately estimate the distances of central stars in PNe. We adopted distances for 50 out of the 124 PNe in our sample, as listed in Table \ref{T:obslog}. Remaining 74 PNe distances are taken from GAIA EDR3\footnote{\url{https://vizier.cds.unistra.fr/viz-bin/VizieR-3?-source=I/352/gedr3dis}}. The distances of these PNe range from 0.3 to 16 kpc. However, the distance of PN G009.8-04.6, reported as 321 pc, is considered uncertain by \cite{2021A&A...656A.110C}. Therefore, we treated the distance as a free parameter in the modeling Under this choice, the remaining parameters vary by $\sim 20\%$, the distance converges to $\approx 4.4,\mathrm{kpc}$, and the value of $\chi^{2}$ decreases to $10.81$.

We assume the ionizing source to be a blackbody to derive T$_{eff}$, following the approach of \cite{1999MNRAS.308..623V} and \cite{2008ApJ...674..954B}. We adopted the T$_{\mathrm{eff}}$ range from \cite{2004A&A...426..779M}, which is given as 60,000–200,000 K. In this study, we set the initial value to 100,000 K and allowed this parameter to vary within the range of 40,000–250,000 K. In our models, the central star SEDs were approximated using blackbody radiation. While this approach is commonly adopted, it tends to overestimate the stellar effective temperature $T_{\mathrm{eff}}$ and luminosity compared to detailed model atmosphere calculations \citep{Bohigas2008, 1999MNRAS.308..623V}. \citet{2000ApJ...543..889G} quantified the typical uncertainties associated with blackbody approximations, finding that $T_{\mathrm{eff}}$ may be overestimated by approximately 10--20\%, with corresponding stellar mass estimates differing by up to 0.05\,$M_{\odot}$.

We assumed the central star's luminosity to be isotropic and kept this parameter variable. To determine the luminosity in our models, we adopted values ranging from 100 \(L_\odot\) to 10$^{5}$ \(L_\odot\), based on the studies of \cite{2008ApJ...674..954B} and \cite{2024A&A...688A..36T}. The iteration steps for both T$_{eff}$ and luminosity were set to 0.01 in logarithmic units.

In a PN geometry, the central star is surrounded by a gas and dust envelope, preventing diffuse radiation from escaping. Due to the small angular size of the selected PNe, their geometry cannot be visually classified. Thus, we assumed a spherical structure, varying the inner radius (R$_{in}$) from 10$^{16}$ cm (0.0032 pc) to 10$^{18}$ cm (0.32 pc) with a 0.05 log-unit step based on \citep{2004A&A...426..779M}. The outer radius was optimized in modeling based on the PN's angular size. In {\scshape Cloudy} modeling, the following parameters;T${eff}$, luminosity, and R${in}$ were treated as variables.

In the radio region, we primarily utilized data at a wavelength of 6 cm. However, since this data was unavailable in the VizieR database, we resorted to using 7.5 cm or 21 cm data. The interstellar extinction is negligible beyond the K-band (2.2 $\mu$m) \citep{2017ApJS..231...22O}, and therefore, we did not use dereddened flux in this region.

The error values associated with the inputs used in the modeling, including fluxes, abundances, IR/radio photometry, and angular sizes, are critical for ensuring the reliability of the results. We used $20{,}000$ iterations in models aimed at minimizing error. Additional iterations did not materially improve performance. The dominant errors come from fluxes (roughly $20\%$).
Emission line fluxes and abundances are accompanied by error values derived from Tan24. For the IRAS and WISE data, a 15\% error margin is assumed \cite{2003AJ....126.1607S}, while the radio fluxes are assigned a 5\% error. Angular sizes are considered to have a 20\% error \citep{2025NewA..11902413Y}. Moreover, fluxes lower than 15\% relative to H$_{\beta}$ are considered unreliable and are excluded from the analysis to ensure the accuracy and validity of the model. These error considerations help to minimize uncertainties and improve the robustness of the model’s predictions. This approximation of the errors is taken from \cite{1999MNRAS.308..623V}. Because the grain-size distribution in our targets is not directly constrained, the grain absorption coefficients in {\scshape cloudy} were computed assuming the canonical MRN grain-size distribution \citep{1977ApJ...217..425M}.

In addition, the other input parameters, $n_H$ and $\rho_d/\rho_g$, used in the model were taken as 500-63,000 and 2.0, respectively, the same values details given in \cite{2024AcA....74..201A}.

Several criteria are employed to halt the modeling process. The one is the temperature which is set to default as there is no ionization below 4,000 K. Additionally, IR fluxes are utilized as one of the criteria for reaching line strength to stop the modeling process. Furthermore, the electron density is set at 0.1 cm$^{-3}$, and a zone count of 1,500 (default 1,400) is adopted as a stopping criterion.

Model inputs generally comprise abundances, densities, and the ionizing star’s T$_{eff}$ and luminosity. Emission lines are not prescribed but computed by the model and then confronted with observations, collectively referred to as the observational constraints. Finally, through iteration, the model outputs the parameter set that minimizes $\chi^{2}$, together with the modelled and observed line intensities.

The parameters are optimized through iterations. While 400 iterations are defined as standard, in our modeling, we opted to adjust the iteration steps to 0.05 and conducted 20,000 iterations for more precise results. This process takes one day for a PN with 8 CPUs and 8 GB RAM. We utilized 56 CPUs and 3,400 GB of RAM on "Hamsi" nodes at the High Performance and Grid Computing Center (TRUBA) in TUBITAK to complete all model computations within a few hours for a PN. As a result of this modeling, using the errors mentioned above, the best-fit values of the parameters are found by calculating the $\chi^{2}$ value. The input and output parameters are given in Table \ref{T:cloudy_in} and Table \ref{T:cloudy_out}, respectively. The SEDs of PNe from the modeling results are given in Fig. \ref{F:model} using pyCloudy. In this model, the input, output, and fitting parameters used are given in Table \ref{model_out}.

\section{Results and Discussions}
\label{sec:res}

In this study, we investigated photoionization models for 124 PN in the Galactic bulge, based on Tan24, to determine the central star's properties and the characteristics of the surrounding gas and dust. 

The photoionization model constructed using {\scshape cloudy}, based on the specified inputs, provides the following results: a T$_{eff}$ range from 50,000 K to 250,000 K, luminosity ranging from 600 L${\odot}$ to 10,000 L${\odot}$. The R${in}$ range from 0.003 to 0.31 pc, defining the location of the radiating region. The $\Theta$ is angular size of a PN which found to vary between 1.7\arcsec - 29\arcsec, which on average is 7\% larger than the observed visible diameter, as the model's $\Theta$ is derived from radio emissions. The derived n${H}$ and $\rho{d}$/$\rho_{g}$ range from 500 to 63,000 cm$^{-3}$ and 2.74 to -0.52, respectively \citep{1999MNRAS.308..623V}.These values obtained from Table \ref{T:cloudy_out} are consistent with previous studies of Galactic PNe, including those in the bulge region (\cite{1999MNRAS.308..623V, 2004A&A...426..779M}; KH22).

The assumption of a homogeneous spherical PN shell with a central star modeled as a blackbody introduces systematic errors. Real PNe are clumpy, stratified, and often asymmetric. As a result, such simplified models commonly underestimate the ionized mass by up to a factor of two, underpredict elemental abundances derived from collisionally excited lines by about 0.2-0.5 dex due to temperature fluctuations, and misestimate ionization correction factors by 20-50\% depending on the geometry and whether the nebula is radiation- or matter-bounded. Consequently, the derived physical conditions and abundances can be significantly distorted. To achieve more realistic results, additional corrections such as including filling factors, adopting multi-zone models, or using non-LTE stellar atmospheres are typically required.

\begin{figure*}
\begin{center}
\begin{tabular}{c@{}c@{}c@{}}
\includegraphics[angle=0,scale=0.27]{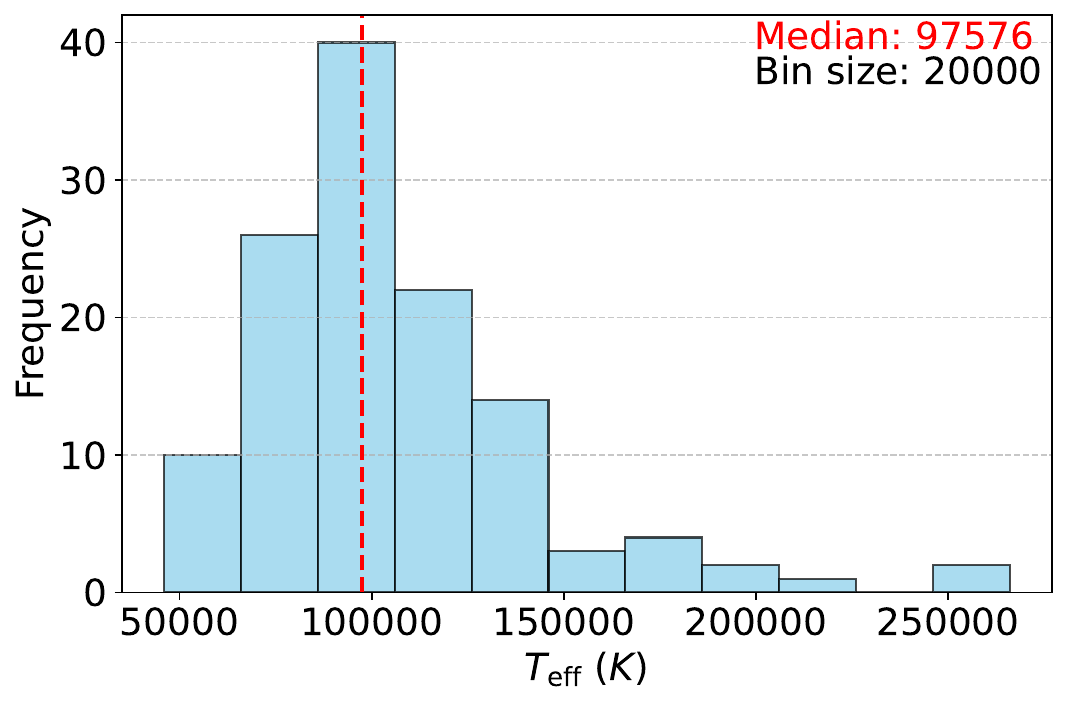} &
\includegraphics[angle=0,scale=0.27]{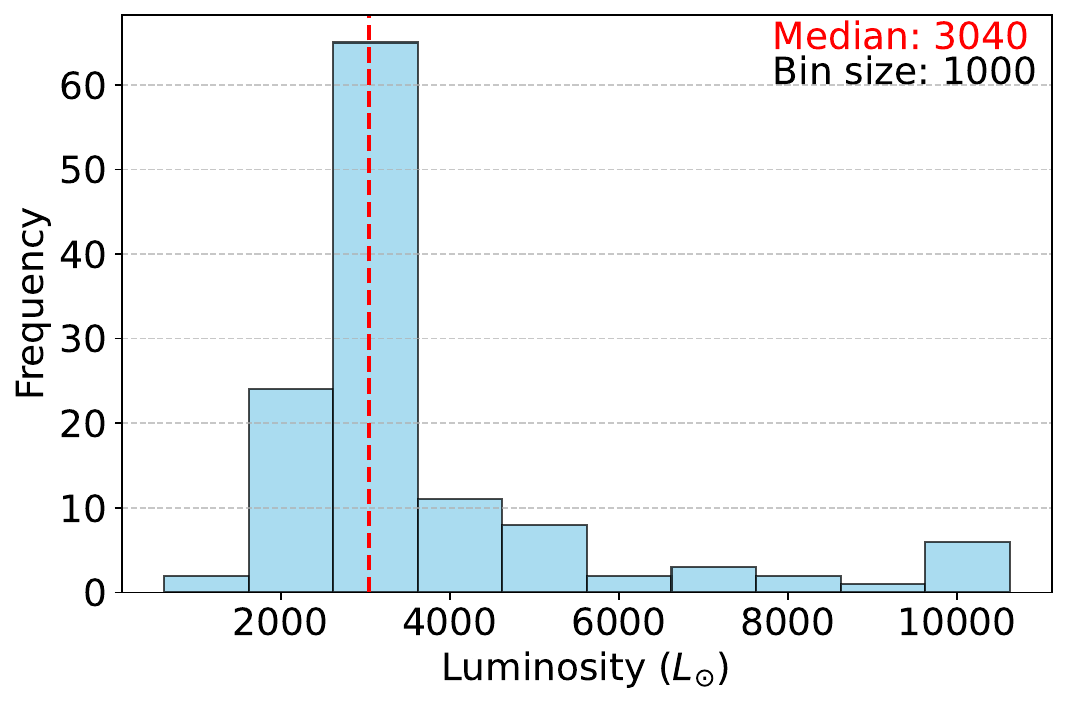} &
\includegraphics[angle=0,scale=0.27]{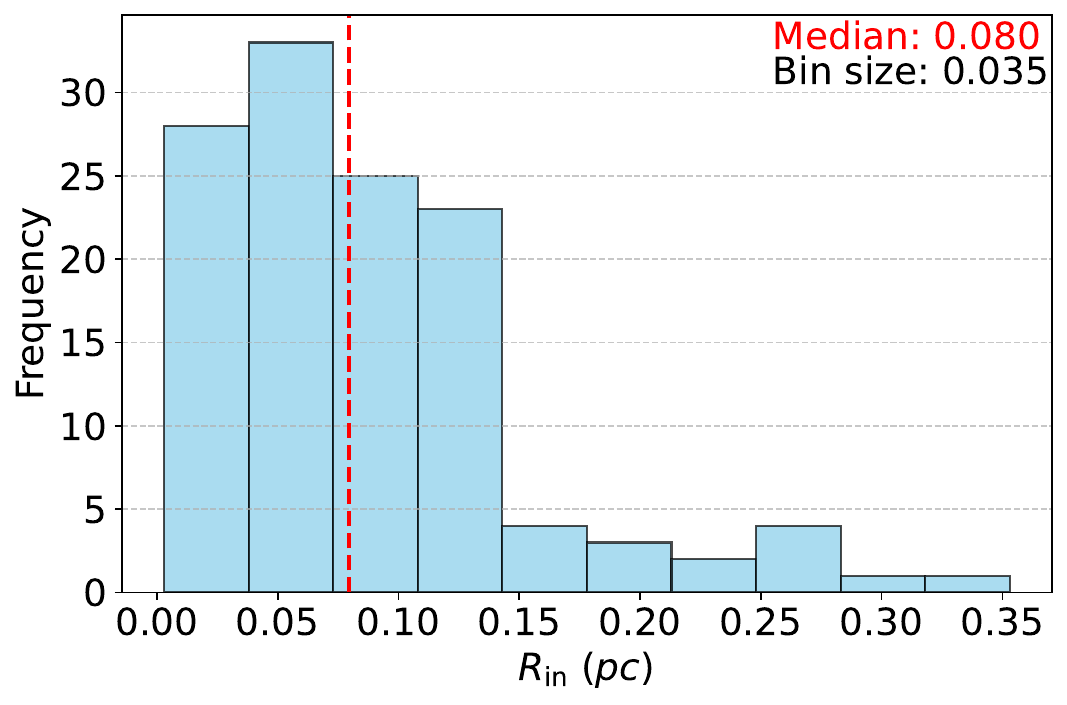} \\
\includegraphics[angle=0,scale=0.27]{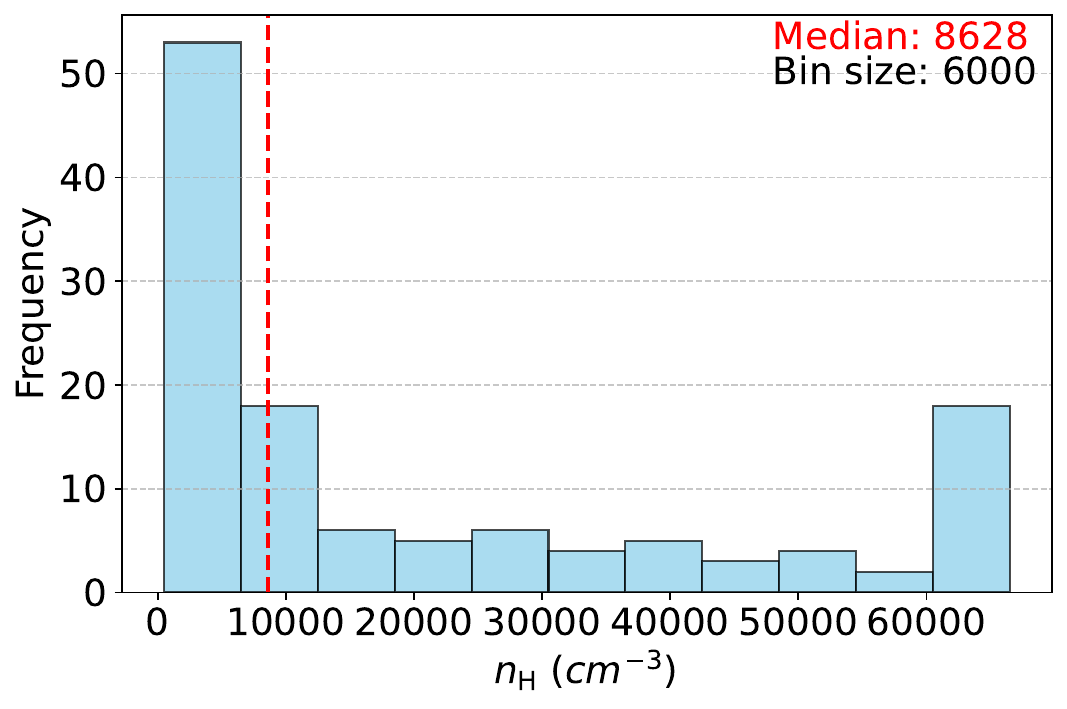} &
\includegraphics[angle=0,scale=0.27]{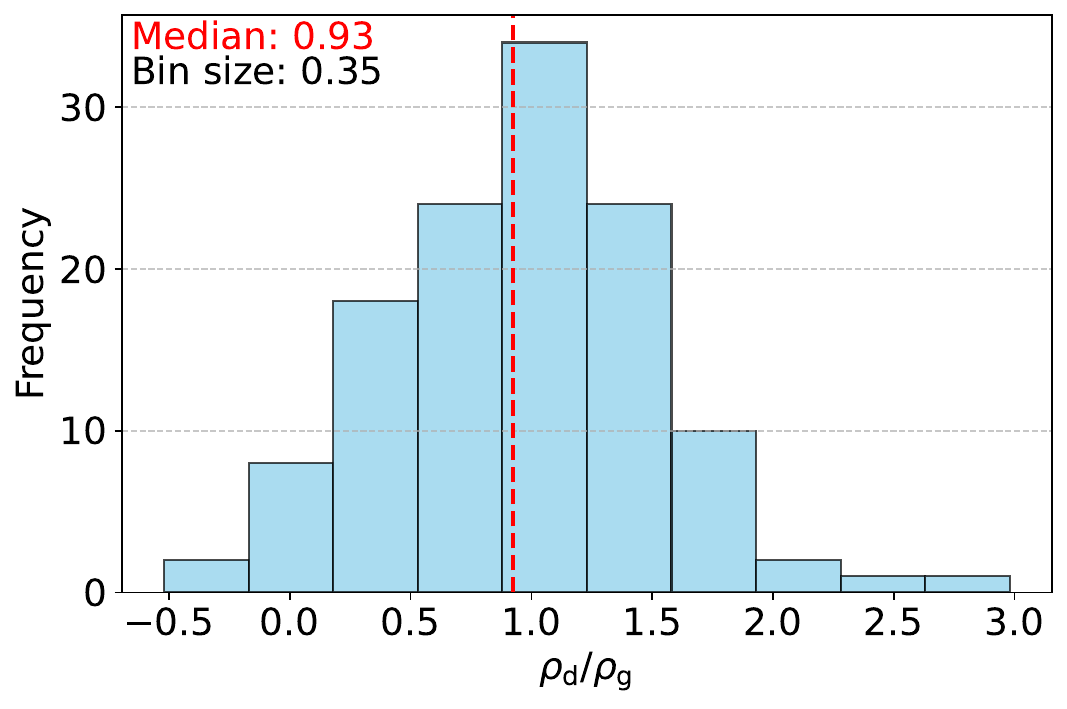} &
\includegraphics[angle=0,scale=0.27]{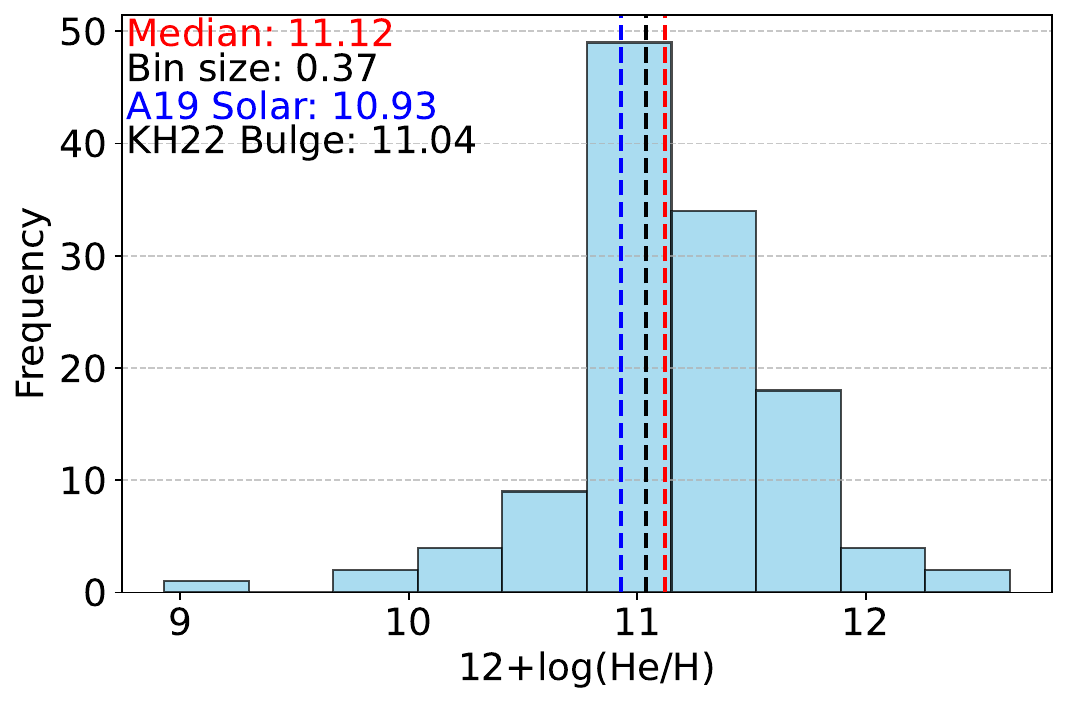} \\
\includegraphics[angle=0,scale=0.27]{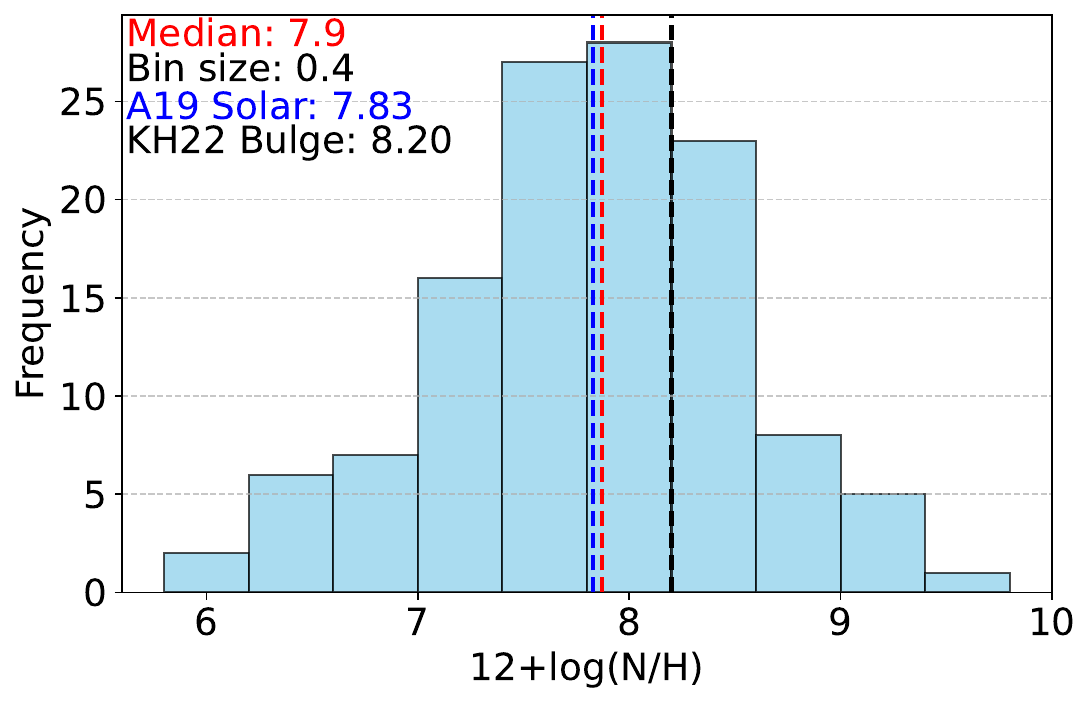} &
\includegraphics[angle=0,scale=0.27]{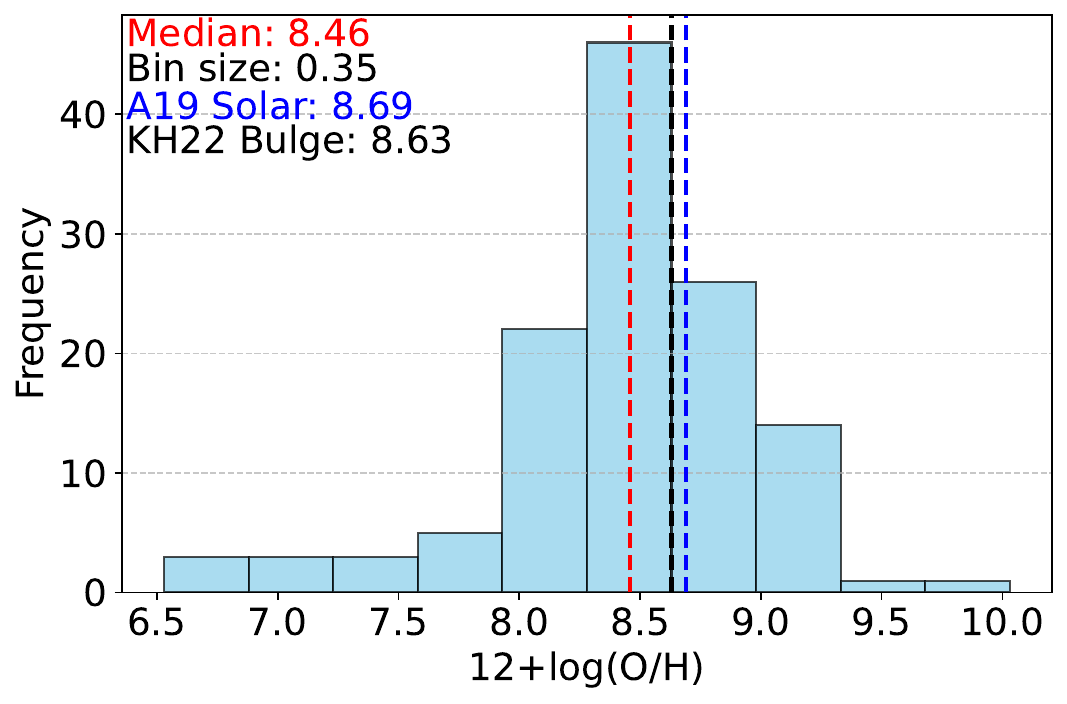} &
\includegraphics[angle=0,scale=0.27]{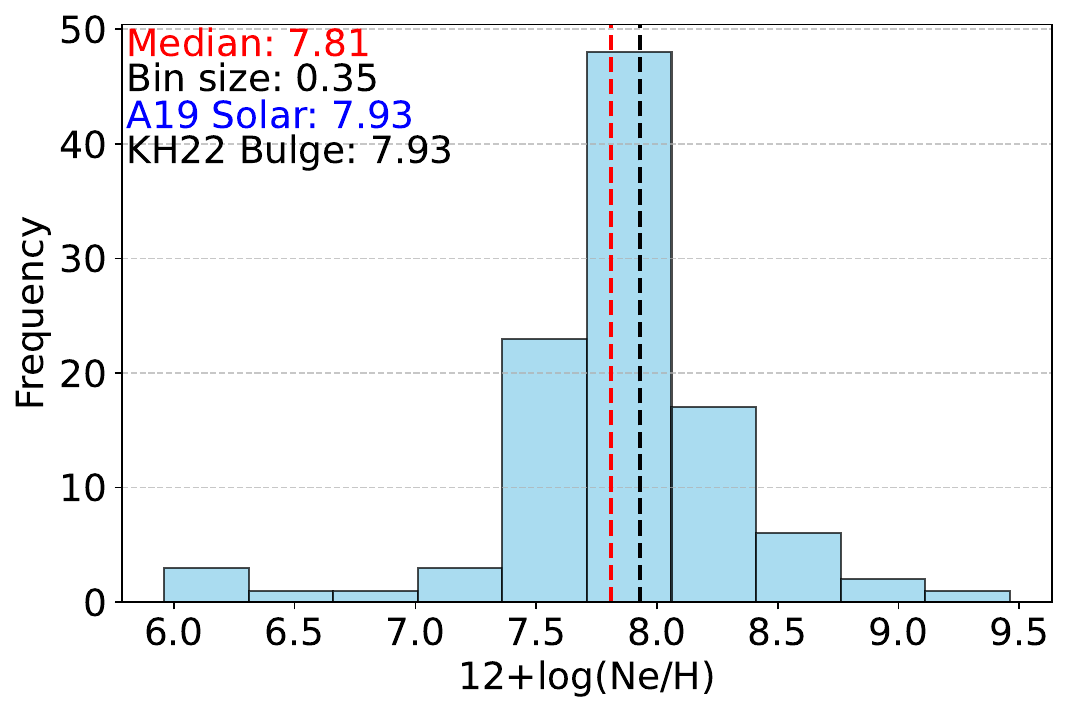} \\
\includegraphics[angle=0,scale=0.27]{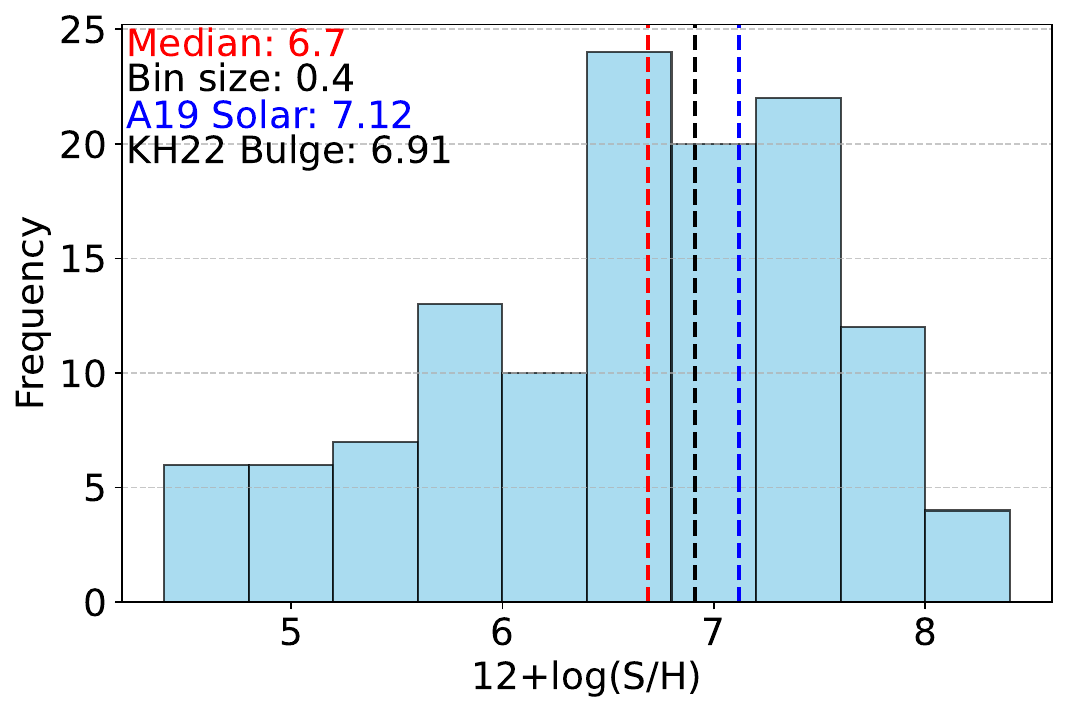} &
\includegraphics[angle=0,scale=0.27]{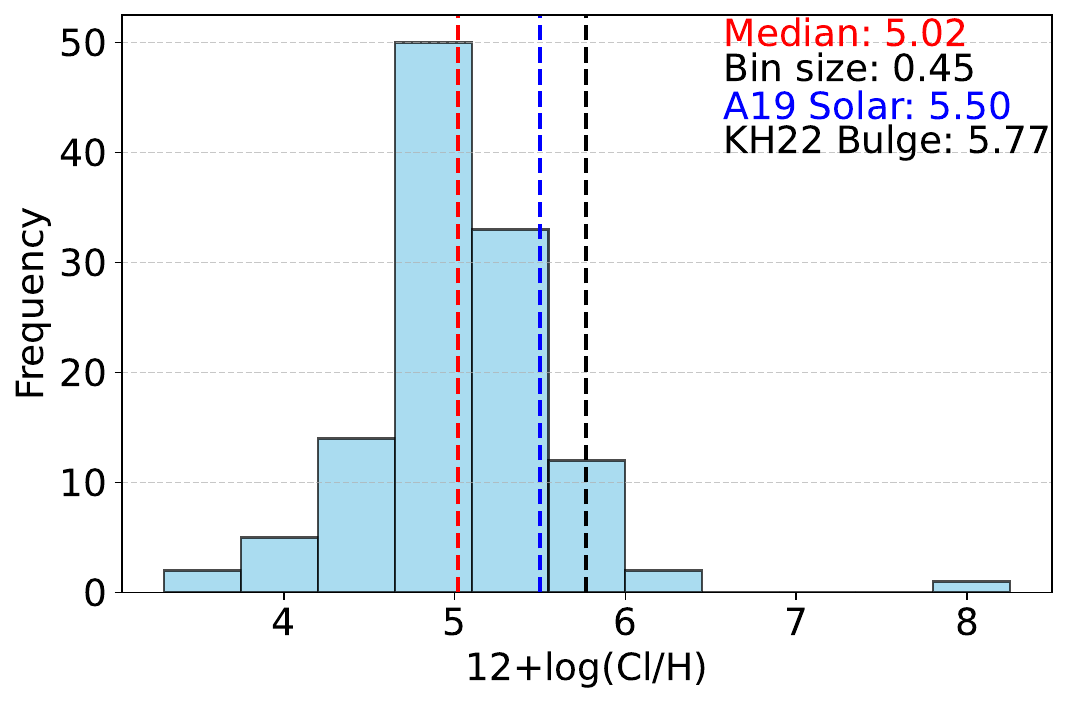} &
\includegraphics[angle=0,scale=0.27]{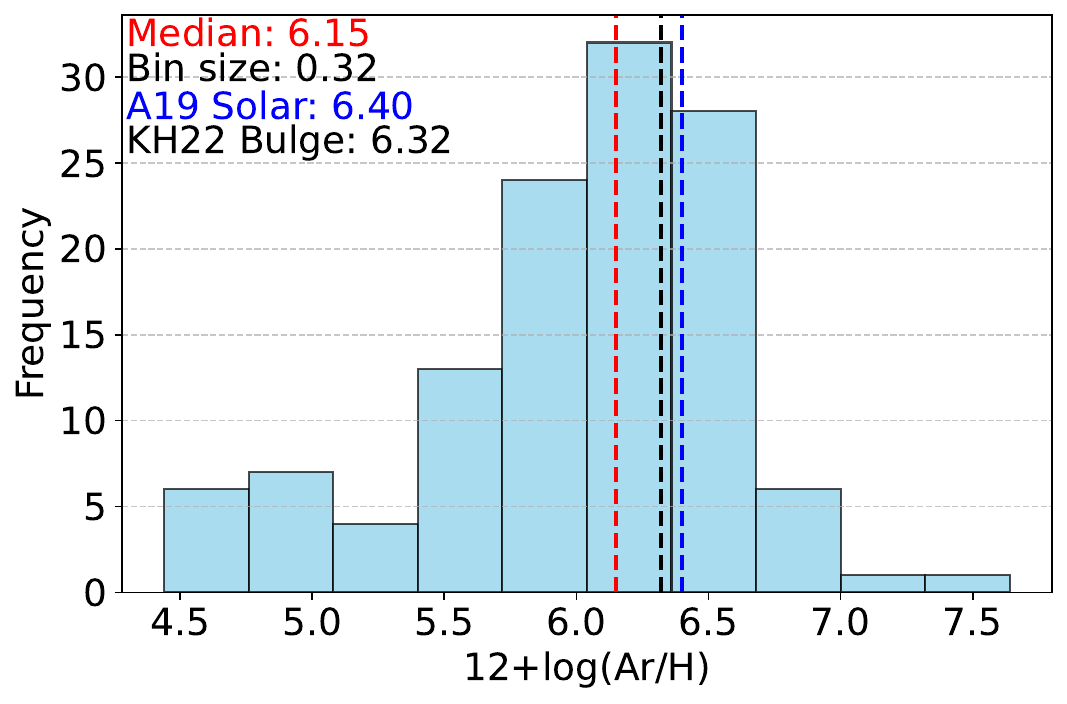}\\

\end{tabular}
\caption{The histograms of parameters output from the {\scshape cloudy} model are presented in Table \ref{T:cloudy_out}. The element abundances are shown in the form of log(X/H) + 12. The red dashed lines indicate the median for each value. The median and bin size are provided at the corresponding location on the plot.} 
\label{F:histograms}
\end{center}
\end{figure*}

We generated histograms to analyze the distribution of each parameter is given in Fig. \ref{F:histograms} using the values from Table \ref{T:cloudy_in}. When the histograms are examined the T$_{eff}$ of the central star distribution peaks at $\sim$ 100,000 K, while luminosities cluster around $\sim$ 3,000 L$_\odot$. The R${in}$ has a median value of 0.08 pc, while n${H}$ is centered around $\sim$ 9,000 cm$^{-3}$, and the ratio of $\rho{d}$/$\rho_{g}$ varies, with a median of 0.9. The median helium abundance (12+log(He/H)) is 11.1, while the abundances of other elements relative to hydrogen are as follows: N/H (7.9), O/H (8.5), Ne/H (7.8), S/H (6.7), Ar/H (6.2) and Cl/H (5.0). In these histograms also display the values from KH22 for the average elemental abundances of PNe in the Galactic bulge and  \cite{2009ARA&A..47..481A} (A19) for the solar abundances. In general, the abundances are relatively consistent within 0.5 dex, with the most notable discrepancy observed in the abundance of Cl/H. S/H and N/H show broader distributions, whereas Cl/H and Ar/H are more tightly clustered.

\begin{figure*}
\begin{center}
\includegraphics[angle=0,scale=0.4]{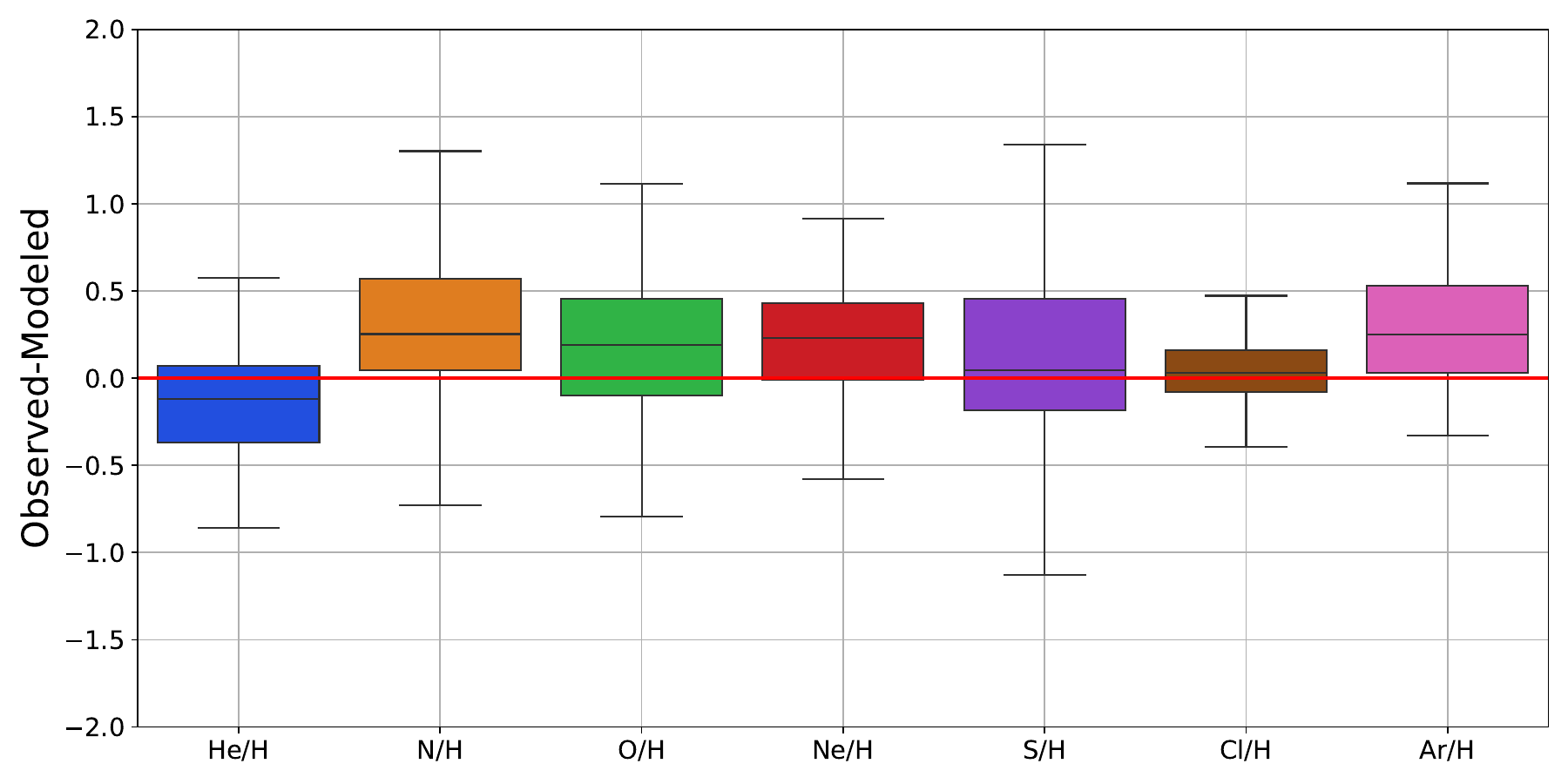}
\caption{Box plot of abundance ratios of He/H, N/H, O/H, Ne/H, S/H, Cl/H and Ar/H between observed data and {\scshape cloudy} model. The y-axis indicates deviations in abundance ratios, with the red line at zero representing no difference between observations and the model. The plot highlights the median, interquartile range, and outliers for each element of PNe.}
\label{F:box}
\end{center}
\end{figure*}

We also compared the elemental abundances determined from the {\scshape cloudy} models with the abundances given by Tan24 (see Fig. \ref{F:box}). The y-axis quantifies the deviations, with each element's plot displaying the median, interquartile range, and outliers to assess model accuracy. The discrepancies between observed and model values were generally less than 0.5 dex. The He/H ratio is lower than the other ratios, while the Cl/H ratio shows a close match between observed and modeled values. However, the S/H values exhibit significant scatter in both the observations and the model.

Known is that the photoelectric heating can significantly influence the thermal balance of ionized nebulae and is best traced through UV fluxes sensitive to grain heating. Very small grains  play a crucial role in this process (\cite{2004MNRAS.350.1330V, 2024AdSpR..74.1366K}. The MRN size distribution \citep{1977ApJ...217..425M}, commonly used in{\scshape cloudy}, lacks sufficient small grains and may therefore underestimate this heating, whereas the KMH distribution \citep{1994ApJ...422..164K} includes these grains and provides a more realistic estimate of photoelectric heating.
In this study, grain heating was not explicitly included because our aim is to model a large, homogeneous sample of 124 PNe and compare their nebular properties. For most objects, the optical emission-line spectra that form the basis of our analysis are only weakly affected by grain heating. While it may be significant in some individual PNe \citep{2024AdSpR..74.1366K}, it is not expected to alter our overall statistical conclusions.

As shown in Table \ref{T:cloudy_in}, the $\chi^{2}$ values, which represent the model's goodness of fit, vary widely with a median value of 44. The chi-squared ($\chi^{2}$) statistic measures how well a model fits the observed data by comparing the differences between observed and expected values relative to their uncertainties. We also calculated ($\chi_{\text{red}}^{2}$) for the models. A low $\chi^{2}$ value indicates that the model and the data are in good agreement, which often suggests higher accuracy. Our {\scshape cloudy} modeling incorporated at least 12 variable parameters, each contributing to the $\chi^{2}$ values. The higher $\chi^{2}$ values were primarily attributed to discrepancies in the IRAS/WISE photometric data. When only optical emission lines were considered in the models, the $\chi^{2}$ values were decreased significantly. Using WISE data instead of IRAS resulted in higher $\chi^{2}$ values, with the highest corresponding to models using WISE data. Therefore, we rely primarily on the IRAS fluxes for lower $\chi^{2}$ values. When the median values were evaluated, they fell within the range of 0.04 to 160, as reported by \cite{1999MNRAS.308..623V} and \cite{2015ApJS..217...22O}.

\begin{figure*}
\begin{center}
\includegraphics[angle=0,scale=0.8]{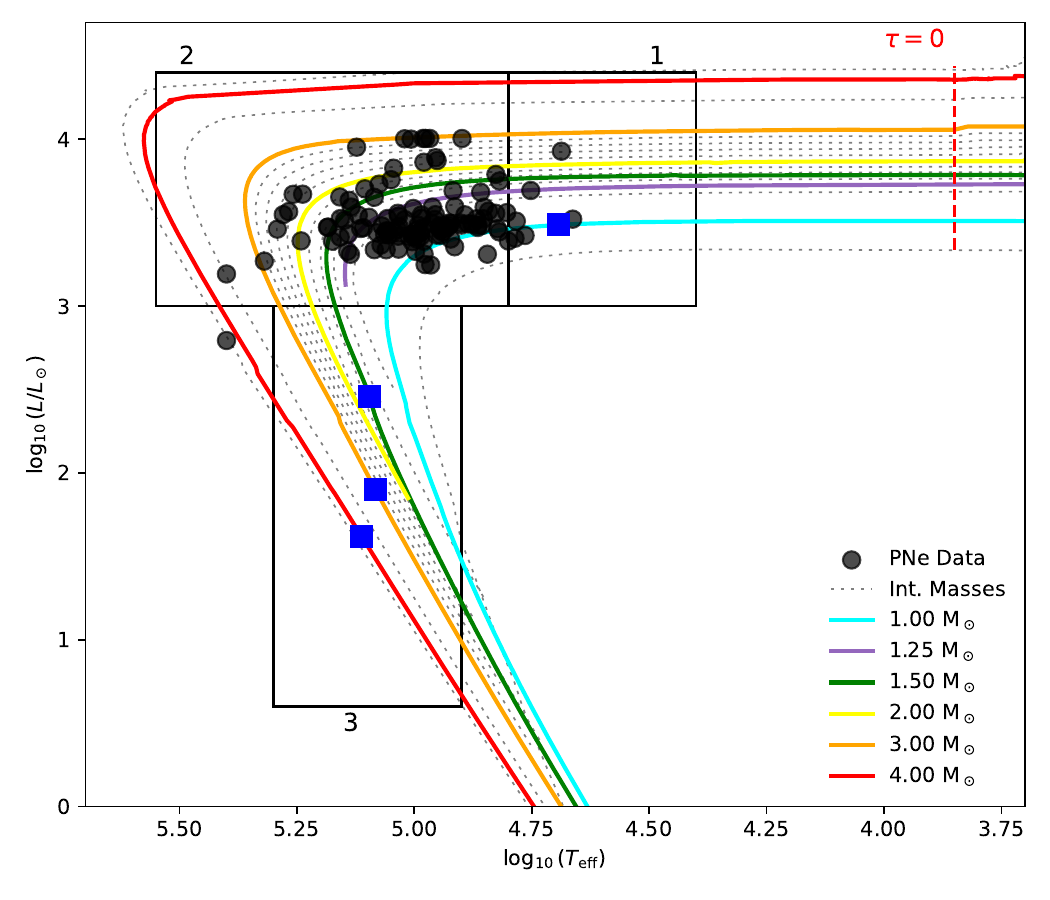}
\caption{The distribution of 124 planetary nebulae (black filled circles) in the H-R diagram is shown using their effective temperatures, $T_{\mathrm{eff}}$, and luminosities, $L/L_{\odot}$. Post-AGB evolutionary tracks are adopted from \citet{2016A&A...588A..25M}. Gray dashed curves denote linearly interpolated tracks at $0.1\,M_{\odot}$ increments (solar metallicity $Z=0.02$). The blue dashed line marks the beginning of the post-AGB phase, defined at $\log T_{\mathrm{eff}}=3.85$ as $\tau=0$. Blue squares indicate ages of $10\,000$~yr measured from $\tau=0$. Black shaded regions are reproduced from \citet{2020Galax...8...29G}.
}
\label{F:cloudy_hr}
\end{center}
\end{figure*}

To estimate the mass and post-AGB age of their CSPN, the Hertzsprung-Russell (H-R) diagram was constructed using the T$_{eff}$ and luminosity values of PNe (see Fig. \ref{F:cloudy_hr}). Evolutionary tracks for post-AGB were taken from \cite{2016A&A...588A..25M}. The evolutionary stage progresses from right to left and then top to bottom on the diagram. More detailed evolutionary tracks were constructed by applying linear interpolation at 0.1\,$M_{\odot}$ intervals, using the existing evolutionary tracks as a basis. \cite{2020Galax...8...29G} divided it into three regions, representing early to late stages, with the nebular mean radius increasing from Region 1 to Region 3 based on the mass and evolutionary age of the CSPN. Most of the 124 PNe fall in Region 2, indicating an intermediate evolutionary stage and a mid-range nebular mean radius.

\begin{figure*}
\begin{center}
\includegraphics[angle=0,scale=0.8]{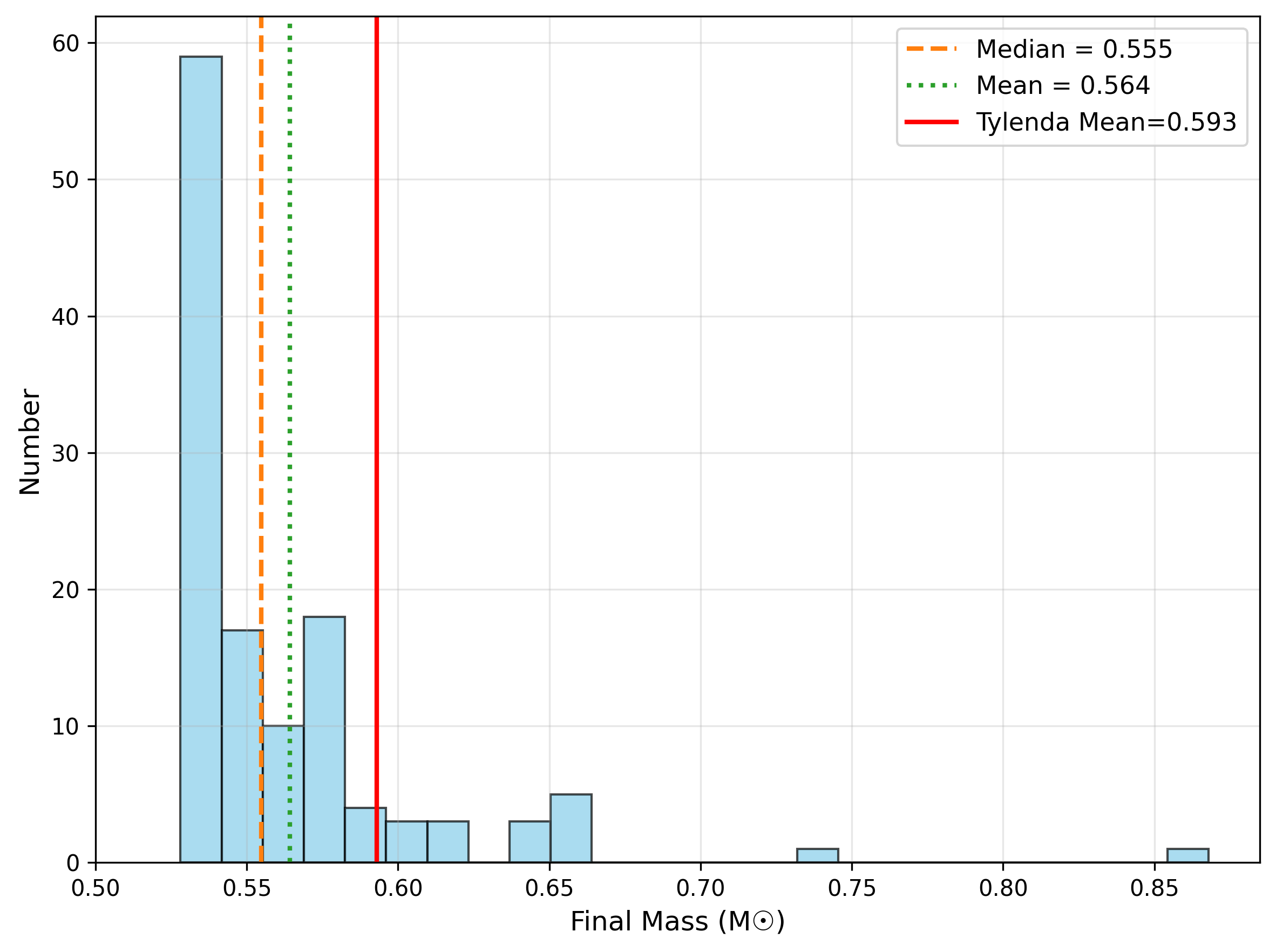}

\caption{This histogram shows final mass distribution of 124 PNe. The x-axis is start with 0.5\(M_\odot\).}
\label{final_mass}
\end{center}
\end{figure*}

The Table \ref{T:hr} provides a detailed summary of the initial and final progenitor masses, mass losses, and evolutionary ages for 124 PNe, derived from Fig. \ref{F:cloudy_hr}. The initial progenitor masses (M$_i$) range from 0.8 to 4.2 \(M_\odot\), with final masses (M$_f$) between 0.53 and 0.87 \(M_\odot\), reflecting significant mass loss in high masses during their evolution. The mass loss values vary from 0.47 to 3.33 \(M_\odot\). The mass-loss rates were computed following the prescriptions of \cite{2016A&A...588A..25M}, where pulsation-enhanced, dust-driven winds are used for the AGB phase (distinguishing O-rich and C-rich regimes), and radiation-driven winds \citep{1995A&A...297..727B} were adopted for the post-AGB phase, both scaled with stellar parameters such as luminosity, metallicity, and pulsation period. \cite{1991A&A...246..221T} reported a mean central star mass of 0.593 \(M_\odot\) for their Galactic bulge PN sample, with a standard deviation of 0.025 \(M_\odot\). In comparison, our study yields a mean value of 0.564 \(M_\odot\) with a standard deviation of 0.045 \(M_\odot\), which lies within the margin of uncertainty and is therefore consistent with their results. The histogram of the final mass of the bulge PNe is given in Fig. \ref{final_mass}. The evolutionary ages $\tau$, measured in years after the beginning of the post-AGB phase, range from 150 years to over 20,000 years, showcasing the wide range of evolutionary stages among these PNe. The age range for Galactic PNe extends up to 30,000 years, as reported in the study by \cite{2022Galax..10...32P}.

\cite{2024ApJ...961L..47T} examined sulfur anomalies in the same Galactic bulge PN population, showing that young, more massive progenitors with distinct dust chemistry display minimal or no sulfur deficit. On the other hand, PNe from intermediate-mass progenitors exhibit more pronounced deficits, likely due to sulfur incorporation into dust grains especially in carbon rich environments which limits its detection in the gas phase. Since our sample corresponds to the lower end of intermediate-mass progenitors, it offers only limited support for the proposed sulfur depletion.

\section{Conclusions}
\label{sec:conc}

This study investigates photoionization models for 124 PNe in the Galactic bulge using {\scshape cloudy} to derive the properties of their central stars and the surrounding gas and dust. The results show:

\begin{itemize}
    \item T$_{\text{eff}}$ range from 50,000 K to 250,000 K, luminosities from 600 L$_{\odot}$ to 10,000 L$_{\odot}$, and R$_{\text{in}}$ between 0.003 and 0.31 pc. The $\Theta$ vary from 1.7" to 29", with an average 7\% larger than the observed visible diameter. n$_H$ range from 500 to 63,000 cm$^{-3}$, and $\rho_{d}$/$\rho_{g}$ range from -0.52 to 2.74.
    \item Histograms of model parameters show that CSPN temperatures peak at $\sim$100,000 K, with luminosities around $\sim$3,000 L$\odot$. The median R${in}$ is 0.08 pc, n$_{H}$ is $\sim$ 9,000 cm$^{-3}$, and the dust-to-gas ratio has a median of 0.9. The elemental abundances are largely consistent within 0.5 dex with KH22 and A19, though the largest discrepancy is seen in Cl/H. Cl/H and Ar/H are more narrowly distributed, whereas S/H and N/H have wider spreads.
    \item We compared the elemental abundances from {\scshape cloudy} models with those in Tan24, finding discrepancies generally within 0.5 dex. The Cl/H ratio from modeling closely matches the observations, while He/H is lower. S/H shows significant scatter in both model and observation.
    \item The $\chi^{2}$ values span a wide range, with a median of 44, consistent with the 0.04–160 range reported in previous studies. Higher $\chi^{2}$ values primarily due to discrepancies in IRAS/WISE photometric data, as using only WISE data significantly worsens $\chi^{2}$. Models incorporating IRAS fluxes generally produce better $\chi^{2}$ values. 
    \item The distribution of elemental abundances is consistent with previous studies, with significant scatter in S/H and N/H, but good agreement between observed and modeled Cl/H.
    \item Most of the 124 PNe are located in Region 2 of the H-R diagram, suggesting an intermediate evolutionary stage and a larger nebular size.
    The initial progenitor masses from 0.8 to 4.2 \(M_\odot\) and final masses between 0.53 and 0.87 \(M_\odot\), indicating significant mass loss (0.47–3.33 \(M_\odot\)) were determined. The derived age range of 150 to over 20,000 years is consistent with the broader Galactic PNe age distribution, which extends up to 30,000 years \citep{2022Galax..10...32P}.
\end{itemize}

Expanding samples, along with improved observations and modeling of the bulge and other Galactic components, will help address unresolved questions about PNe.

\section*{Acknowledgements}

This research was supported by the Scientific and Technological Research Council of Turkey (TÜBİTAK) through project number 122F122. We also thank TÜBİTAK National Observatory (TUG) staff. NA, thanks to Peter A. Maria van Hoof for his contributions to the process of {\scshape cloudy} modeling. This work has made use of data from the European Space Agency (ESA) mission Gaia (www.cosmos.esa.int/gaia), processed by the Gaia Data Processing and Analysis Consortium (DPAC, www.cosmos.esa.int/web/gaia/dpac/consortium). Funding for the DPAC has been provided by national institutions, in particular, the institutions participating in the Gaia Multilateral Agreement. The numerical calculations reported in this paper were fully/partially performed at TUBITAK ULAKBIM, High Performance and Grid Computing Center (TRUBA resources). During the preparation of this work the author(s) used ChatGPT-3.5 in order to English wording of certain phrases. After using this tool/service, the author(s) reviewed and edited the content as needed and take(s) full responsibility for the content of the publication.

\paragraph{Supplementary Material}
\label{sec:Supl}

\begin{figure*}[t]
\centering
\setlength{\tabcolsep}{20pt} 
\renewcommand{\arraystretch}{1} 

\resizebox{0.65\textwidth}{!}{%
\begin{tabular}{c@{}c@{~~}c@{}}
  \textbf{ PN G002.8+01.7} & & \textbf{ PN G352.1+05.1} \\
  \includegraphics[width=0.48\linewidth]{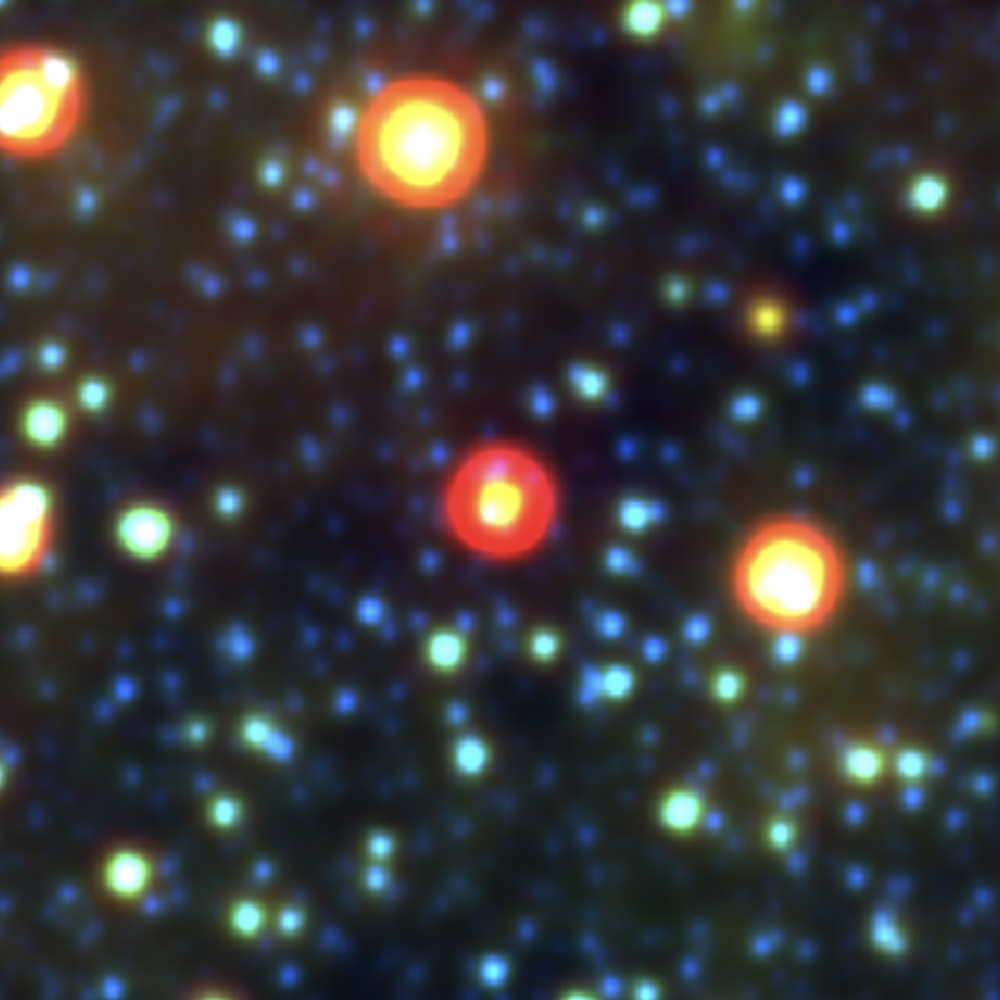} & &
  \includegraphics[width=0.48\linewidth]{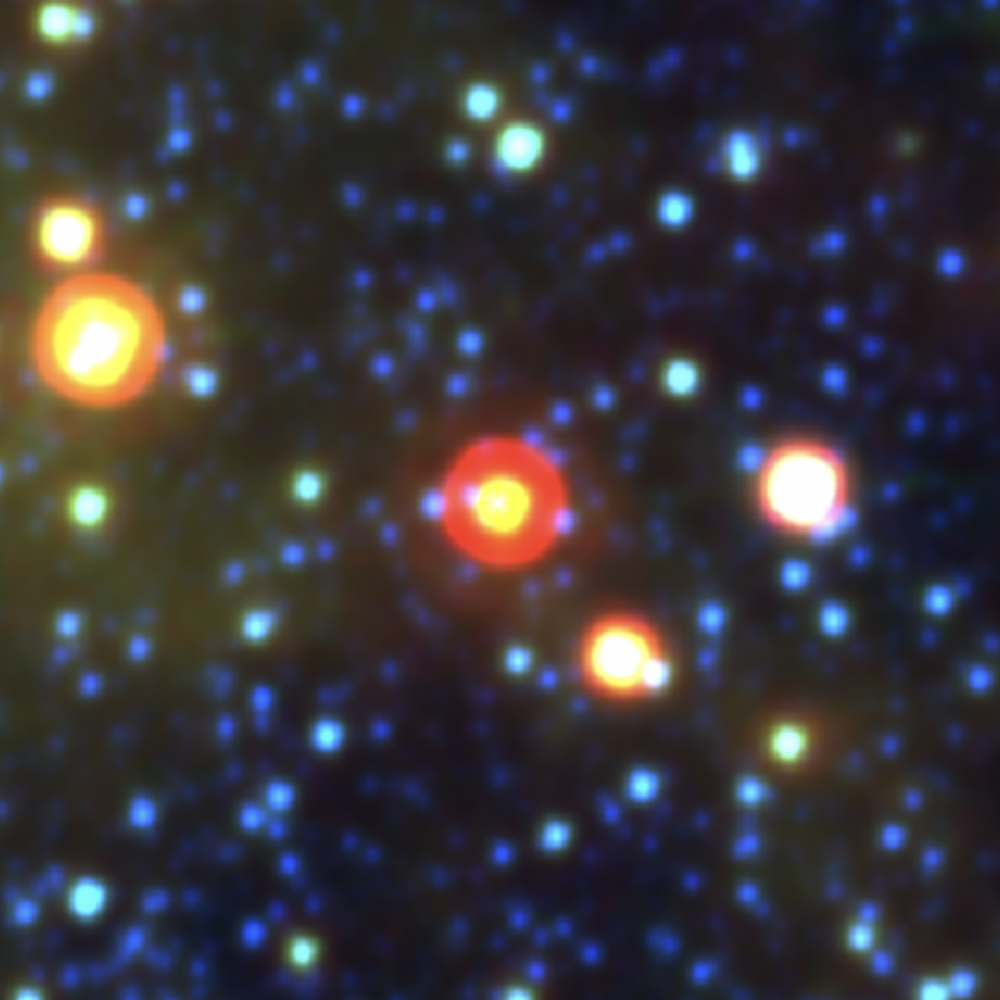} \\
  \textbf{ PN G007.6+06.9} & &\textbf{ PN G359.3$-$01.8} \\
  \includegraphics[width=0.48\linewidth]{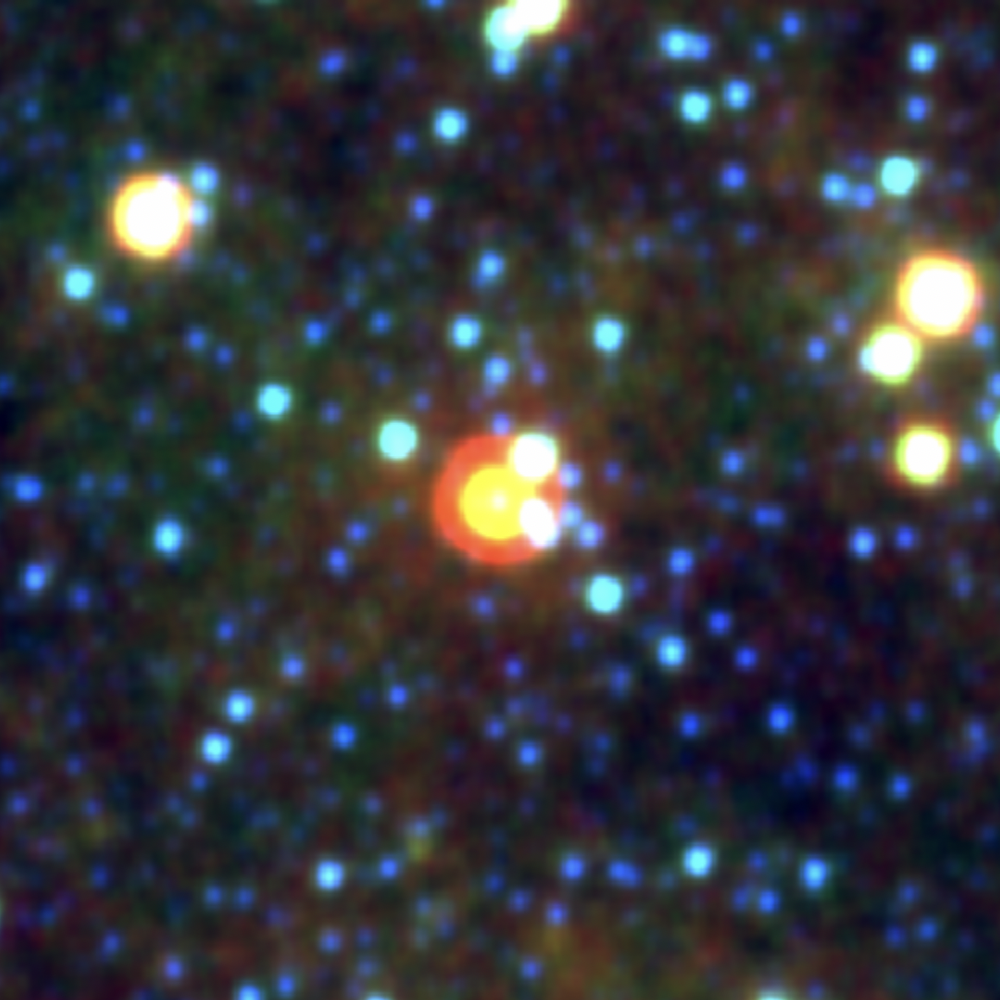} & &
  \includegraphics[width=0.48\linewidth]{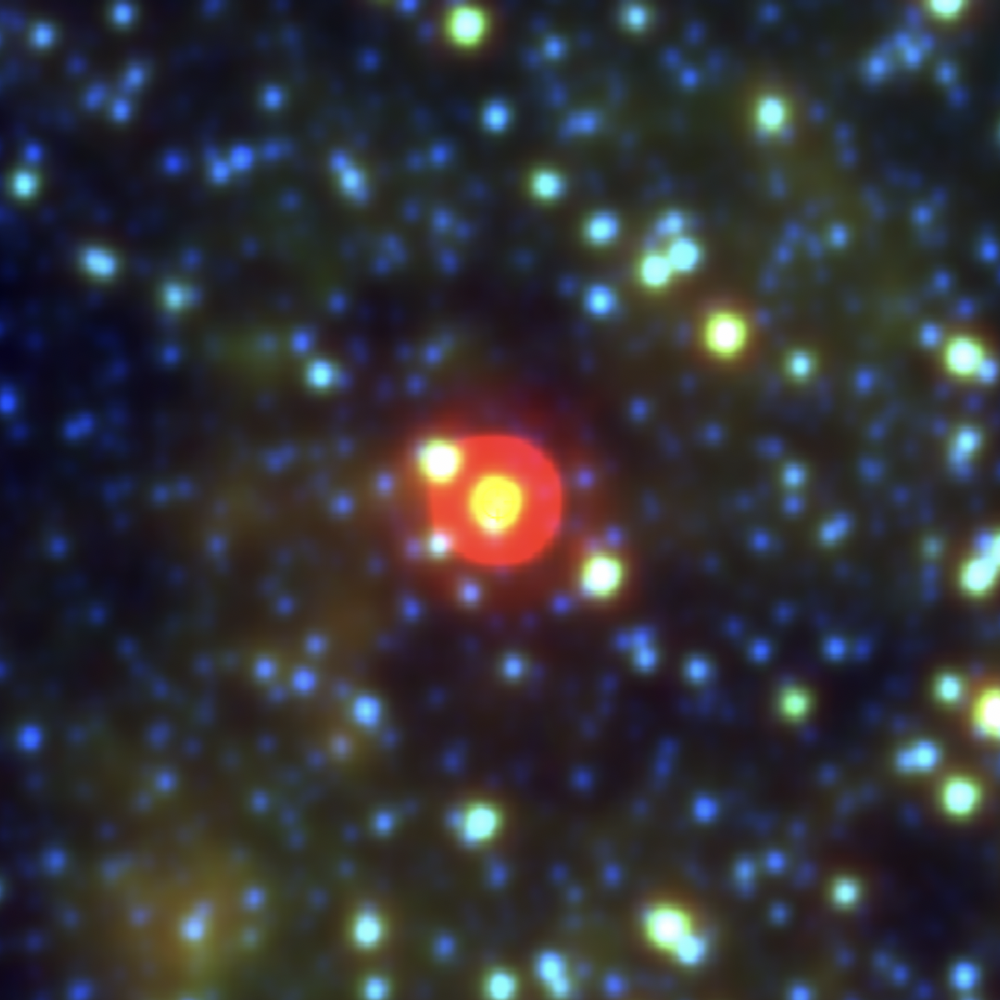}
\end{tabular}%
}

\caption{RGB composite images of PN\,G002.8+01.7, PN\,G352.1+05.1, PN\,G007.6+06.9, and PN\,G359.3$-$01.8 generated using WISE data.  
The images are composed with Red: W4, Green: W3, and Blue: W2 channels, and are arranged according to the PN\,G designations listed in Table~\ref{T:obslog}.  
Each planetary nebula is located at the center of its panel, typically exhibiting a reddish hue.  
Each image spans $500\arcsec \times 500\arcsec$, with north up and east to the left. Remaining PNe images have given as supplementary data.}
\label{fig:wise}
\end{figure*}

\texttt{WISE\_Bulge\_PNe.zip}: Complete version of Fig.~\ref{fig:wise}, containing RGB composite images of PNe generated from WISE data. These supplementary files related to this article will be made available online.

\bibliographystyle{plainnat} 
\bibliography{pne}

\begin{table*}

\centering
\footnotesize
\caption{Source list of 124 PNe. The table includes the PN G (Planetary Nebula Galactic coordinates) designations, common names, right ascension (R.A.) and declination (Dec.) in J2000 coordinates, angular diameters (in arcseconds), R$_{mag}$ magnitudes, and distances (in parsecs). Distances in bold face are taken from the catalog of \protect\cite{2021A&A...656A.110C}.}
    \begin{tabular}{l@{}c@{}c@{}c@{}c@{}c@{}r|l@{}c@{}c@{}c@{}c@{}c@{}r}
    \hline\hline
    PN G & Name & R.A (J2000) & Dec. (J2000) & Diam. & R$_{mag}$ & Distance &PN G & Name & R.A (J2000) & Dec. (J2000) & Diam. & R$_{mag}$ & Distance \\
    &&(hh:mm:ss.s)& (dd:mm:ss.s) & (\arcsec) &  & (pc)&    &&(hh:mm:ss.s)& (dd:mm:ss.s) & (\arcsec) &  & (pc)  \\
\hline
000.1-02.3 &  Bl 3-10   & 17 55 20.5 & -29 57 36.1 &   7.2    &        17.91         & 6696 & 007.8-04.4 & H 1-65 & 18 20 08.8 & -24 15 05.3 & 8 & 13.7        & \textbf{9930}  \\
000.1+02.6 & Al 2-J      & 17 35 35.5 & -27 24 06.5 &   10.5     &                 & 7401  & 008.2+06.8 & SaSt 2-15  & 17 38 57.4 & -18 17 35.9 & 1.8  & 13.7        & \textbf{14902}  \\
000.1+04.3 & H 1-16      & 17 29 23.4 & -26 26 05.8 & 5.0  & 15.9 & 6633                   & 008.4-03.6 & H 1-64      & 18 18 23.9 & -23 24 57.5 & 7.6  & 17.1        & 5700  \\
000.2-01.9 & M 2-19      & 17 53 45.6 & -29 43 47.0 & 20.0 & 14.4          & \textbf{6490}          & 008.6-02.6 & MaC 1-11    & 18 14 51.0 & -22 43 55.7  & 8.4  & 16.7       & 6007  \\
000.2-04.6 & WRAY 16-363 & 18 04 44.0 & -31 02 47.4  &    6.4    &                 & 6853  & 009.4-09.8 & M 3-32      & 18 44 43.0 & -25 21 34.8  & 8.1  & 15.3        & \textbf{8099}  \\
000.3-04.6 & M 2-28      & 18 05 02.7 & -30 58 18.0  &    5.5    &                 & 6208  & 009.8-04.6 & H 1-67      & 18 25 05.0 & -22 34 52.6 & 7.0  & 16       & \textbf{321}   \\
000.3+06.9 & Terz N 41 & 17 20 22.0 & -24 51 52.2 & 15.2 & 19.1          & 6786            & 351.1+04.8 & M 1-19      & 17 03 46.8 & -33 29 44.4 &   8.0     & 14.9         & \textbf{7078}  \\
000.4-01.9 & M 2-20      & 17 54 25.3 & -29 36 08.2  & 6.0  & 16.6          & 10475        & 351.2+05.2 & M 2-5       & 17 02 19.1 & -33 10 04.8 &  6.5      &               & \textbf{7192}  \\
000.4-02.9 & M 3-19      & 17 58 19.4 & -30 00 39.3  &    7.2    &                 & \textbf{4629}  & 351.6-06.2 & H 1-37      & 17 50 44.6 & -39 17 25.8  &  8.6      &               & 6169  \\
000.7-02.7 & M 2-21      & 17 58 09.6 & -29 44 20.1  & 2.8  & 14.9          & 8340          & 351.9-01.9 & WRAY 16-286 & 17 33 00.7 & -36 43 51.8  &   10.0  &               & \textbf{7487}  \\
000.7-07.4 & M 2-35      & 18 17 37.2 & -31 56 46.8  &   5.0     &                 & 7822  & 351.9+09.0 & PC 13       & 16 50 17.1 & -30 19 55.5  &  10.0      &               & 6753  \\
000.7+03.2 & M 4-5       & 17 34 54.7 & -26 35 57.3  & 6.7  & 17.5          & 5959         & 352.0-04.6 & H 1-30      & 17 45 06.8 & -38 08 49.7  &      5.4  &               & 5610  \\
000.9-02.0 & Bl 3-13     & 17 56 02.8 & -29 11 16.6 & 4.2  & 16.5          & \textbf{7511}          & 352.1+05.1 & M 2-8       & 17 05 30.7 & -32 32 08.1  &   5.0     &               & 3675  \\
000.9-04.8 & M 3-23      & 18 07 06.2 & -30 34 17.8  &     13.6   &        & \textbf{5374}  & 352.6+03.0 & H 1-8       & 17 14 42.9 & -33 24 47.4  &      3.4  &               & 6425  \\
001.1-01.6 & Sa 3-92     & 17 54 52.1 & -28 48 55.3 & 6.4  & 17.7          & \textbf{10094}         & 353.2-05.2 & H 1-38      & 17 50 45.2 & -37 23 52.1  &   14.0     &               & 5132  \\
001.2-03.0 & H 1-47      & 18 00 37.6 & -29 21 50.6 & 2.5  & 14.5          & \textbf{8922}          & 353.3+06.3 & M 2-6       & 17 04 18.3 & -30 53 28.9 &   8.0     & 15.7        & 7353  \\
001.2+02.1 & Hen 2-262  & 17 40 12.8 & -26 44 21.9  & 4.6  & 16.9          & \textbf{5923}          & 353.7+06.3 & M 2-7       & 17 05 13.9 & -30 32 19.7  &     7.8   &               & 5826  \\
001.3-01.2 & Bl M        & 17 53 47.1 & -28 27 17.1 & 4.0  & 17.4          & \textbf{4842}          & 354.5+03.3 & Th 3-4      & 17 18 51.9 & -31 39 06.6 &  14.0      &               & 8340  \\
001.4+05.3 & H 1-15      & 17 28 37.6 & -24 51 07.2 & 5.0  & 15.3          & 5771          & 354.9+03.5 & Th 3-6      & 17 19 20.2 & -31 12 40.8 &        &               & 5620  \\
001.6-01.3 & Bl Q        & 17 54 35.0 & -28 12 43.6 & 3.8  & 16.7          & 2499          & 355.4-02.4 & M 3-14      & 17 44 20.6 & -34 06 40.6  &     8.0   &               & 1222  \\
001.7-04.4 & H 1-55      & 18 07 14.6 & -29 41 24.7 & 3.0  & 15          & \textbf{9683}            & 355.9-04.2 & M 1-30      & 17 52 59.0 & -34 38 22.6  &   3.5     &               & \textbf{6275}  \\
001.7+05.7 & H 1-14      & 17 28 01.8 & -24 25 23.3 & 6.6  & 17.3          & 6850          & 355.9+03.6 & H 1-9       & 17 21 31.9 & -30 20 48.6 &     5.0   & 14.8        & \textbf{6119}  \\
002.0-06.2 & M 2-33      & 18 15 06.6 & -30 15 33.3 &    5.4    & 14.6           & \textbf{5003}    & 356.1-03.3 & H 2-26      & 17 49 50.8 & -34 00 30.5  &     5.5   &               & \textbf{15775} \\
002.1-02.2 & M 3-20      & 17 59 19.4 & -28 13 48.2  & 6.6  & 15.4          & 7429         & 356.3-06.2 & M 3-49      & 18 02 32.0& -35 13 14.0  &    9.8    &               & 6148  \\
002.1-04.2 & H 1-54      & 18 07 07.3 & -29 13 06.0 & 1.9  & 13.6          & \textbf{7001}          & 356.5-03.6 & H 2-27      & 17 51 50.6 & -33 47 36.3 &     5.2   &               & \textbf{8817} \\
002.2-09.4 & Cn 1-5      & 18 29 11.7 & -31 29 59.1 &    7.2    & 13.3           & \textbf{4574}    & 356.8-05.4 & H 2-35      & 18 00 18.3 & -34 27 39.3 &     7.0   &               & 8340  \\
002.3+02.2 & K 5-11      & 17 42 30.1 & -25 45 28.7  &     12.0   & 14.5           & 4740  & 356.8+03.3 & Th 3-12     & 17 25 06.1 & -29 45 17.0 & 2.0  & 16.3        & \textbf{16151}  \\
002.5-01.7 & Pe 2-11     & 17 58 31.3 & -27 37 05.8  & 7.8  & 17.8          & 6562         & 356.9+04.4 & M 3-38      & 17 21 04.5 & -29 02 59.7 & 1.6  & 15.8        & 6989  \\
002.6+02.1 & Terz N 1580 & 17 43 39.4 & -25 36 42.5 & 11.7 & 18.1          & 6288          & 357.0+02.4 & M 4-4       & 17 28 50.3 & -30 07 45.0 &  6.3      &               & 9313  \\
002.7-04.8 & M 1-42      & 18 11 05.0 & -28 58 59.1  & 13.1 & 14.8          & \textbf{4307}         & 357.1-04.7 & H 1-43      & 17 58 14.4 & -33 47 37.6 &   2.0     &               & \textbf{1819}  \\
002.8+01.7 & H 2-20      & 17 45 39.8 & -25 40 00.3  & 2.8  & 16.1          & \textbf{11847}         & 357.1+03.6 & M 3-7       & 17 24 34.5 & -29 24 19.8 & 6.5    & 15.3        & \textbf{5669}  \\
002.8+01.8 & Terz N 1567 & 17 45 28.3 & -25 38 10.4 &    11.8    &                 & 6719  & 357.1+04.4 & Terz N 18   & 17 21 38.0 & -28 55 14.9 &   10.9     &               & 6374  \\
002.9-03.9 & H 2-39      & 18 08 05.8 & -28 26 10.5 & 6.9  & 17.0          & \textbf{10326}          & 357.2+02.0 & H 2-13      & 17 31 08.1 & -30 10 28.3  & 5.6 & 19.0        & 5593  \\
003.2-06.2 & M 2-36      & 18 17 41.4 & -29 08 19.9  & 8.1  & 14.0          & 1034         & 357.3+04.0 & H 2-7       & 17 23 24.9 & -28 59 06.1 & 5.7  & 17.3        & \textbf{7491}  \\
003.6-02.3 & M 2-26      & 18 03 11.9 & -26 58 31.2  & 10.5 & 17.6          & 6388         & 357.5+03.1 & Th 3-16     & 17 27 24.4 & -29 21 14.5 & 6.0  & 15.1        & 7476  \\
003.7-04.6 & M 2-30      & 18 12 34.4 & -27 58 10.5  & 5.1  & 14.5          & 5989         & 357.5+03.2 & M 3-42      & 17 26 59.9 & -29 15 32.1  & 7.2  & 17.8 & 5935  \\
003.7+07.9 & H 2-8       & 17 24 45.8 & -21 33 35.8 &  11.9      &                 & 8340  & 357.6-03.3 & H 2-29      & 17 53 16.9 & -32 40 38.5 &     10.7   &               & \textbf{5368}  \\
003.8-04.3 & H 1-59      & 18 11 29.3 & -27 46 15.7 & 6.6  & 15.7          & 7377          & 357.9-03.8 & H 2-30      & 17 56 14.2 & -32 37 19.5 & 13.3       &             & 6734  \\
003.9-02.3 & M 1-35      & 18 03 39.3 & -26 43 33.9  & 7.3  & 15.3          & 6320         & 357.9-05.1 & M 1-34      & 18 01 22.1 & -33 17 45.2 &     29.0   &               & 6333  \\
003.9-03.1 & KFL 7       & 18 06 50.0 & -27 06 16.1 &    8.1 &                 & 2388      & 358.0+09.3 & Th 3-1      & 17 05 44.5 & -25 25 01.7 & 10.0 & 17.0        & 7564  \\
004.0-03.0 & M 2-29      & 18 06 40.9 & -26 54 56.4 & 4.8  & 14.7 & \textbf{9694}      & 358.2+03.5 & H 2-10      & 17 27 32.9 & -28 31 06.9  & 3.7  & 16.3        & \textbf{8976}  \\
004.1-03.8 & KFL 11      & 18 10 12.2 & -27 16 35.0 & 3.0  & 17.2          & 5853          & 358.2+04.2 & M 3-8       & 17 24 52.2 & -28 05 54.6  & 5.0  & 15.8        & \textbf{6223}  \\
004.2-03.2 & KFL 10      & 18 08 01.3 & -26 54 01.5 & 7.1  & 17          & \textbf{14375}            & 358.5-04.2 & H 1-46      & 17 59 02.5 & -32 21 43.5 &     3.0   &               & \textbf{7895}  \\
004.2-04.3 & H 1-60      & 18 12 25.2 & -27 29 12.8  & 6.0  & 15.7          & \textbf{6100}         & 358.5+02.9 & Al 2-F      & 17 30 30.4 & -28 35 54.9  & 4.2  & 18.3        & \textbf{10801}  \\
004.6+06.0 & H 1-24      & 17 33 37.6 & -21 46 24.8 & 9.0  & 16          & \textbf{6409}            & 358.6-05.5 & M 3-51      & 18 04 56.2 & -32 54 02.0  &     20.9   &               & \textbf{5328}  \\
004.8-05.0 & M 3-26      & 18 16 11.4 & -27 14 58.3  & 11.0 & 16.7          & \textbf{7553}         & 358.6+07.8 & M 3-36      & 17 12 39.7 & -25 43 37.5 & 4.2  & 16.0        & 8838  \\
004.8+02.0 & H 2-25      & 17 49 00.5 & -23 42 54.6 & 3.1  & 15.9          & \textbf{8855}          & 358.7+05.2 & M 3-40      & 17 22 28.3 & -27 08 42.4 & 2.2  & 15.9        & 6868  \\
005.0-03.9 & H 2-42      & 18 12 23.3 & -26 32 54.0 &    13.0    &                 & 6235  & 358.8+03.0 & Th 3-26     & 17 31 09.3 & -28 14 50.4  & 9.1  & 18.3        & 7249  \\
005.2+05.6 & M 3-12      & 17 36 22.6 & -21 31 12.2 & 6.0  & 16.8          & 6539          & 358.9+03.4 & H 1-19      & 17 30 02.6 & -27 59 17.5  & 2.6  & 15.5        & 9123  \\
005.5-04.0 & H 2-44      & 18 13 40.5 & -26 08 38.3  &    6.6    &                 & 7678  & 359.0-04.1 & M 3-48      & 17 59 56.8 & -31 54 28.1  &  5.4      &               & 7899  \\
005.5+06.1 & M 3-11      & 17 35 21.4 & -20 57 23.4 & 7.2  & 15.9          & 7252          & 359.1-02.9 & M 3-46      & 17 55 05.7 & -31 12 16.4  &    15.5    &               & 6970  \\
005.8-06.1 & NGC 6620    & 18 22 54.2 & -26 49 17.2 & 7.4  &14.3 & 6939                    & 359.2+04.7 & Th 3-14     & 17 25 44.1 & -26 57 47.9 & 1.7  & 15.1        & \textbf{13451}  \\
006.1+08.3 & M 1-20      & 17 28 57.6 & -19 15 54.0 & 2.5  & 14          & \textbf{7617}            & 359.3-01.8 & M 3-44      & 17 51 18.9 & -30 23 53.2 &    4.4    &               & \textbf{6922}  \\
006.4-04.6 & Pe 2-13     & 18 18 13.4 & -25 38 10.3 & 8.0  & 15.4          & 6265          & 359.6-04.8 & H 2-36      & 18 04 07.8 & -31 39 11.4  &    17.7    &               & \textbf{6930}  \\
006.4+02.0 & M 1-31      & 17 52 41.4 & -22 21 57.2 & 3.5  & 14.2          & 4974          & 359.7-01.8 & M 3-45      & 17 52 06.0 & -30 05 14.0  &   7.1     &               & \textbf{5164}  \\
006.8-03.4 & H 2-45      & 18 14 28.8 & -24 43 38.4&    4.6    &                 & 5950    & 359.8-07.2 & M 2-32      & 18 14 50.6 & -32 36 55.2 &    8.0    &               & \textbf{9501}  \\
006.8+02.3 & Th 4-7      & 17 52 22.6 & -21 51 13.4 & 4.0  & 16.9 & 2004                   & 359.8+02.4 & Th 3-33     & 17 35 48.1 & -27 43 20.4 & 6.0  & 16.6        & 8652  \\
007.0-06.8 & Vy 2-1      & 18 27 59.6 & -26 06 48.1 & 4.0  & 13.9          & \textbf{7243}          & 359.8+03.7 & Th 3-25     & 17 30 46.8 & -27 05 58.0  & 3.0  & 16.6        & 8055  \\
007.0+06.3 & M 1-24      & 17 38 11.6 & -19 37 37.6 &    6.4    &                 & 4761   & 359.8+05.2 & Terz N 19   & 17 25 23.6 & -26 11 52.9 & 19.0 & 18.6        & 7533  \\
007.5+07.4 & M 1-22      & 17 35 10.2 & -18 34 20.4 & 9.0  & 18.1          & \textbf{6527}          & 359.8+05.6 & M 2-12      & 17 24 01.5 & -25 59 23.3 & 4.4  & 13.6        & 5101  \\
007.6+06.9 & M 1-23      & 17 37 22.0 & -18 46 41.8  & 7.0  & 16.8          & 6428         & 359.8+06.9 & M 3-37      & 17 19 13.4 & -25 17 17.6 &   10.0     &               & 6241  \\
007.8-03.7 & M 2-34      & 18 17 15.9 & -23 58 53.7  & 8.0  & 15.6          & 6044         & 359.9-04.5 & M 2-27      & 18 03 52.6 & -31 17 47.2 &    3.3    &               & 6799 \\

\hline \hline

    \end{tabular}
\label{T:obslog}
\end{table*}
\begin{table*}
\centering
\caption{Photoionization emission lines obtained from Tan24.}
\begin{tabular}{ll ll ll ll}
\hline
Line & $\lambda$ (\AA) & Line & $\lambda$ (\AA) & Line & $\lambda$ (\AA) & Line & $\lambda$ (\AA) \\
\hline
$[$O\,\textsc{ii}$]$ & 3726 & $[$Ar\,\textsc{iv}$]$ & 4740 & He\,\textsc{i} & 5876 & $[$S\,\textsc{ii}$]$ & 6731 \\
$[$O\,\textsc{ii}$]$ & 3729 & H$\beta$  & 4861 & $[$S\,\textsc{iii}$]$ & 6312 & $[$Ar\,\textsc{v}$]$ & 7006 \\
$[$Ne\,\textsc{iii}$]$ & 3869 & $[$O\,\textsc{iii}$]$ & 4959 & $[$Ar\,\textsc{v}$]$ & 6436 & $[$Ar\,\textsc{iii}$]$ & 7136 \\
H$\delta$  & 4102 & $[$O\,\textsc{iii}$]$ & 5007 & $[$N\,\textsc{ii}$]$ & 6548 & $[$O\,\textsc{ii}$]$ & 7319 \\
H$\gamma$  & 4340 & He\,\textsc{ii} & 5412 & H$\alpha$  & 6563 & $[$O\,\textsc{ii}$]$ & 7330 \\
He\,\textsc{i} & 4472 & $[$Cl\,\textsc{iii}$]$ & 5518 & $[$N\,\textsc{ii}$]$ & 6584 & $[$Cl\,\textsc{iv}$]$ & 7531 \\
He\,\textsc{ii} & 4686 & $[$Cl\,\textsc{iii}$]$ & 5538 & He\,\textsc{i} & 6678 & $[$Ar\,\textsc{iii}$]$ & 7751 \\
$[$Ar\,\textsc{iv}$]$ & 4711 & $[$N\,\textsc{ii}$]$ (auroral) & 5755 & $[$S\,\textsc{ii}$]$ & 6716 & $[$Cl\,\textsc{iv}$]$ & 8046 \\
\hline
\end{tabular}
\label{flux_table}
\end{table*}

\begin{table}
\centering
\caption{Wavelength, observed and modeled, $\chi^{2}$ of emission lines and Photometry/Radio values of PN G356.8+03.3 (see Fig. \ref{F:model}).}
\begin{tabular}{lc@{}c@{~}c@{}c@{}}
\toprule
Observable & $\lambda$ & Observed & Modeled    & $\chi^{2}$ \\
\midrule
\multicolumn{5}{l}{\textit{Emission lines}}\\
H$\delta$                  & 4101.73 & 0.280  & 0.272  & $2.67\times10^{-1}$ \\
H$\gamma$                  & 4340.46 & 0.479  & 0.474  & $5.82\times10^{-2}$ \\
H$\beta$                   & 4861.32 & 1.009  & 1.009  & $9.81\times10^{-7}$ \\
$[${O\,\textsc{iii}}$]$     & 5006.84 & 0.264  & 0.263  & $5.73\times10^{-3}$ \\
$[$N\,\textsc{ii}$]$       & 6548.05 & 0.996  & 1.039  & $1.17$ \\
H$\alpha$                  & 6562.80 & 2.872  & 2.876  & $2.13\times10^{-3}$ \\
$[$N\,\textsc{ii}$]$       & 6583.45 & 2.935  & 2.760  & $1.61$ \\
\midrule
\multicolumn{5}{l}{\textit{Photometry / Radio (Jy)}}\\
$F_{\nu}$                  & $12\,\mu\mathrm{m}$     & 0.026  & 2.45     & - \\
$F_{\nu}$                  & $25\,\mu\mathrm{m}$     & 2.344    & 2.25     & $7.75\times10^{-2}$ \\
$F_{\nu}$                  & $60\,\mu\mathrm{m}$     & 5.545    & 5.89     & $1.72\times10^{-1}$ \\
$F_{\nu}$                  & $100\,\mu\mathrm{m}$    & 2.762    & 25.44    & - \\
$S_{\nu}$                  & $6\,\mathrm{cm}$        & 0.005 & 0.005   & $2.03\times10^{-5}$ \\
\midrule
Angular diameter           & (arcsec)                & 2        & 2.085    & $4.49\times10^{-2}$ \\
\bottomrule

\end{tabular}
\label{model_out}
\end{table}

\onecolumn
\scriptsize
\begin{longtable}{l@{~~}c@{}c@{}c@{}c@{}c@{}c@{}c@{}c@{}c@{}c@{}c@{}c@{}c@{}c@{}c@{}c@{}c@{}c@{}c@{}c@{}c@{}c@{}c@{}c@{}r@{}r@{}}
\caption{Modeled and observed values for 124 PNe. The "<" symbol represents the upper limit. The table includes IR and radio intensities (5GHz: J/ApJ/714/1096/table1,
4GHz: J/ApJS/255/30/comp,
1.4GHz: J/ApJS/117/361/table1 and J/MNRAS/474/5008/spidxcat), angular diameters, the chi-square ($\chi^2$ ) and the reduced chi-square, ($\chi_{\text{red}}^{2}$). All WISE and radio fluxes have been multiplied by 100 for consistency. Observed values are presented on the right, while model values are shown on the left for each flux column.}
\label{T:cloudy_in}\\
\hline
PN G & \multicolumn{2}{c}{IRAS12} & \multicolumn{2}{c}{IRAS25} & \multicolumn{2}{c}{IRAS60} & \multicolumn{2}{c}{IRAS100} & \multicolumn{2}{c}{WISE1} & \multicolumn{2}{c}{WISE2} & \multicolumn{2}{c}{WISE3} & \multicolumn{2}{c}{WISE4} & \multicolumn{2}{c}{5 GHz} & \multicolumn{2}{c}{4 GHz} & \multicolumn{2}{c}{1.4 GHz} & \multicolumn{2}{c}{$\Theta$}  & $\chi^{2}$ & $\chi_{\text{red}}^{2}$ 
\\
& \multicolumn{2}{c}{(12 $\mu$m)} & \multicolumn{2}{c}{(25 $\mu$m)} & \multicolumn{2}{c}{(60 $\mu$m)} & \multicolumn{2}{c}{(100 $\mu$m)} & \multicolumn{2}{c}{(3.35 $\mu$m)} & \multicolumn{2}{c}{(4.6 $\mu$m)} & \multicolumn{2}{c}{(11.6 $\mu$m)} & \multicolumn{2}{c}{(22.1 $\mu$m)} & \multicolumn{2}{c}{(6 cm)} & \multicolumn{2}{c}{(7.5 cm)} & \multicolumn{2}{c}{(21.4 cm)} & \multicolumn{2}{c}{(arcsec)} & \\
\hline
\endfirsthead

\multicolumn{26}{c}%
{{\bfseries \tablename\ \thetable{} -- continued}} \\
\hline
PN G & \multicolumn{2}{c}{IRAS12} & \multicolumn{2}{c}{IRAS25} & \multicolumn{2}{c}{IRAS60} & \multicolumn{2}{c}{IRAS100} & \multicolumn{2}{c}{WISE1} & \multicolumn{2}{c}{WISE2} & \multicolumn{2}{c}{WISE3} & \multicolumn{2}{c}{WISE4} & \multicolumn{2}{c}{5 GHz} & \multicolumn{2}{c}{4 GHz} & \multicolumn{2}{c}{1.4 GHz} & \multicolumn{2}{c}{$\Theta$} & $\chi^{2}$ & $\chi_{\text{red}}^{2}$ \\
& \multicolumn{2}{c}{(12 $\mu$m)} & \multicolumn{2}{c}{(25 $\mu$m)} & \multicolumn{2}{c}{(60 $\mu$m)} & \multicolumn{2}{c}{(100 $\mu$m)} & \multicolumn{2}{c}{(3.35 $\mu$m)} & \multicolumn{2}{c}{(4.6 $\mu$m)} & \multicolumn{2}{c}{(11.6 $\mu$m)} & \multicolumn{2}{c}{(22.1 $\mu$m)} & \multicolumn{2}{c}{(6 cm)} & \multicolumn{2}{c}{(7.5 cm)} & \multicolumn{2}{c}{(21.4 cm)} & \multicolumn{2}{c}{(arcsec)} & \\
\hline
\endhead

\hline
\endfoot
000.1-02.3 &  &  &  &  &  &  &  &  & 0.73 & 10.90 & 0.77 & 7.59 & 2.52 & 11.80 & 35.39 & 32.50 &  &  & 3.24 & 0.95 &  &  & 6.2 & 5.5 & 5541 & 2770.4 \\
000.1+02.6 &  &  &  &  &  &  &  &  & 0.34 & 2.71 & 0.35 & 1.68 & 0.74 & 3.94 & 44.20 & 38.70 &  &  & 1.48 & 0.68 &  &  & 6.3 & 10.5 & 1508 & 376.9 \\
000.1+04.3 &  &  & 2.1 & <0.7 & 5.1 & 5.1 & 1.8 & <10.3 &  &  &  &  &  &  &  &  & 5.83 & 5.84 &  &  &  &  & 7.1 & 5.0 & 96 & 48.2 \\
000.2-01.9 & 0.005 & <2.2 & 0.6 & 1 & 3 & 2.8 & 1.5 & <59.2 &  &  &  &  &  &  &  &  & 1.43 & 1.40 &  &  &  &  & 12.2 & 20.0 & 21 & 10.6 \\
000.2-04.6 &  &  &  &  &  &  &  &  & 0.10 & 0.27 & 0.11 & 0.21 & 0.94 & 3.23 & 26.60 & 8.26 &  &  & 0.46 & 0.45 &  &  & 6.6 & 6.4 & 559 & 79.9 \\
000.3-04.6 & 0.01 & <2.3 & 1.1 & 0.8 & 2.9 & 2.9 & 1.1 & <52.1 &  &  &  &  &  &  &  &  & 0.97 & 1.00 &  &  &  &  & 9.9 & 5.5 & 46 & 5.7 \\
000.3+06.9 &  &  &  &  &  &  &  &  & 0.04 & 0.01 & 0.04 & 0.02 & 0.05 & 0.24 & 1.79 & 0.45 &  &  &  &  & 0.19 & 0.25 & 18.8 & 15.2 & 711 & 118.5 \\
000.4-01.9 & 0.01 & <2.7 & 2.4 & 3.8 & 6.4 & 6.2 & 2.4 & <186 &  &  &  &  &  &  &  &  & 0.32 & 0.32 &  &  &  &  & 3.9 & 6.0 & 12 & 4.1 \\
000.4-02.9 & 0.01 & <4.4 & 2 & 1.5 & 5.1 & 5.1 & 1.9 & <63 &  &  &  &  &  &  &  &  & 0.59 & 0.60 &  &  &  &  & 12.8 & 7.2 & 18 & 6.0 \\
000.7-02.7 & 0.3 & 1 & 5.5 & 1.5 & 8.0 & <7.8 & 3.1 & <99.1 &  &  &  &  &  &  &  &  & 2.30 & 2.30 &  &  &  &  & 1.3 & 2.8 & 171 & 170.7 \\
000.7-07.4 &  &  &  &  &  &  &  &  & 0.06 & 0.10 & 0.06 & 0.12 & 0.76 & 3.03 & 23.80 & 5.96 & 0.27 & 0.46 &  &  &  &  & 5.2 & 5.0 & 423 & 84.6 \\
000.7+03.2 & 0.01 & <2 & 1.7 & 2 & 4.5 & 4.5 & 1.7 & <16.1 &  &  &  &  &  &  &  &  & 1.49 & 1.50 &  &  &  &  & 9.2 & 6.7 & 20 & 3.4 \\
000.9-02.0 &  &  &  &  &  &  &  &  & 0.24 & 0.56 & 0.26 & 0.51 & 6.06 & 14.90 & 177 & 135 & 1.10 & 1.00 &  &  &  &  & 3.6 & 4.2 & 60 & 59.7 \\
000.9-04.8 & 0.3 & 0.5 & 6.1 & 3.3 & 4 & 4 & 1.0 & <38.8 &  &  &  &  &  &  &  &  & 2.79 & 2.80 &  &  &  &  & 7.5 & 13.6 & 33 & 16.5 \\
001.1-01.6 &  &  &  &  &  &  &  &  & 0.28 & 0.88 & 0.30 & 0.53 & 0.36 & 4.45 & 12.80 & 11.40 &  &  & 1.29 & 0.51 &  &  & 10.4 & 6.4 & 2483 & 413.9 \\
001.2-03.0 & 0.1 & <1.3 & 5.2 & 5.5 & 7 & 7 & 2.2 & <83.8 &  &  &  &  &  &  &  &  & 1.00 & 1.00 &  &  &  &  & 2.5 & 2.5 & - & - \\
001.2+02.1 &  &  &  &  &  &  &  &  & 0.55 & 1.21 & 0.58 & 1.22 & 6.80 & 21.40 & 159 & 133 & 2.43 & 2.11 &  &  &  &  & 3.0 & 4.6 & 107 & 53.3 \\
001.3-01.2 & 0.1 & <3.3 & 4.7 & 3.2 & 5.5 & 5.5 & 1.6 & <334 &  &  &  &  &  &  &  &  & 1.69 & 1.70 &  &  &  &  & 5.4 & 4.0 & 10 & 4.8 \\
001.4+05.3 & 0.03 & <0.3 & 2.5 & 2.7 & 3.5 & 2.8 & 1.1 & <33.9 &  &  &  &  &  &  &  &  & 1.30 & 1.30 &  &  &  &  & 6.3 & 5.0 & 10 & 10.4 \\
001.6-01.3 & 0.02 & <3.4 & 5 & 3.5 & 42 & 15.4 & 34.5 & 111 &  &  &  &  &  &  &  &  &  &  & 2.62 & 2.63 &  &  & 7.1 & 3.8 & 113 & 28.2 \\
001.7-04.4 & 0.01 & <1.4 & 1.6 & 1.2 & 3.8 & 3.8 & 1.4 & <42.7 &  &  &  &  &  &  &  &  & 0.53 & 0.53 &  &  &  &  & 4.1 & 3.0 & - & - \\
001.7+05.7 & 0.02 & <0.4 & 0.8 & 0.7 & 1 & 1 & 0.3 & <37 &  &  &  &  &  &  &  &  & 2.19 & 2.20 &  &  &  &  & 7.2 & 6.6 & 5 & 2.7 \\
002.0-06.2 & 0.04 & <0.4 & 2.6 & 2.3 & 3 & 3 & 0.9 & <3.4 &  &  &  &  &  &  &  &  & 2.14 & 2.15 &  &  &  &  & 6.8 & 5.4 & 7 & 7.4 \\
002.1-02.2 &  &  &  &  &  &  &  &  & 0.84 & 0.38 & 0.90 & 0.61 & 6.88 & 14.80 & 176 & 158 & 3.77 & 4.00 &  &  &  &  & 3.1 & 6.6 & 79 & 79.1 \\
002.1-04.2 & 0.8 & <2.4 & 5.1 & 4.8 & 1.6 & 1.6 & 0.3 & <64.2 &  &  &  &  &  &  &  &  & 2.91 & 2.94 &  &  &  &  & 2.3 & 1.9 & 70 & 35.1 \\
002.2-09.4 & 1.2 & 1.7 & 14.2 & 7.5 & 19.2 & 9.4 & 6.7 & 6.7 &  &  &  &  &  &  &  &  & 4.36 & 4.40 &  &  &  &  & 9.1 & 7.2 & 27 & 5.3 \\
002.3+02.2 &  &  &  &  &  &  &  &  & 0.16 & 0.30 & 0.17 & 0.27 & 2.51 & 10.40 & 81.73 & 20.40 &  &  & 0.73 & 1.22 &  &  & 10.3 & 12.0 & 540 & 108.1 \\
002.5-01.7 &  &  &  &  &  &  &  &  & 0.10 & 0.10 & 0.11 & 0.09 & 1.16 & 3.38 & 32.63 & 8.23 &  &  & 0.46 & 0.61 &  &  & 6.3 & 7.8 & 267 & 29.7 \\
002.6+02.1 &  &  &  &  &  &  &  &  & 0.26 & 1.03 & 0.28 & 1.25 & 2.59 & 15.40 & 149 & 113 &  &  & 1.17 & 0.88 &  &  & 4.4 & 11.7 & 740 & 105.7 \\
002.7-04.8 & 0.01 & <1.2 & 2 & 0.9 & 14.4 & 7.1 & 7.6 & 7.6 &  &  &  &  &  &  &  &  & 2.32 & 2.40 &  &  &  &  & 19.5 & 13.1 & 33 & 8.4 \\
002.8+01.7 &  &  &  &  &  &  &  &  & 0.71 & 4.93 & 0.76 & 4.24 & 2.87 & 6.74 & 128 & 128 & 3.23 & 1.63 &  &  &  &  & 4.2 & 2.8 & 1040 & 346.6 \\
002.8+01.8 &  &  &  &  &  &  &  &  & 0.10 & 0.20 & 0.11 & 0.20 & 0.62 & 6.00 & 23.37 & 11.90 &  &  & 0.48 & 1.32 &  &  & 7.3 & 11.8 & 2296 & 328.0 \\
002.9-03.9 &  &  &  &  &  &  &  &  & 0.19 & 0.16 & 0.20 & 0.23 & 0.77 & 2.42 & 40.96 & 34.60 &  &  & 0.85 & 0.80 &  &  & 3.7 & 6.9 & 90 & 44.9 \\
003.2-06.2 & 0.2 & <0.3 & 6.2 & 2.3 & 5 & 5 & 1.3 & <9.6 &  &  &  &  &  &  &  &  & 2.38 & 2.50 &  &  &  &  & 17.9 & 8.1 & 83 & 13.8 \\
003.6-02.3 &  &  &  &  &  &  &  &  & 0.08 & 0.17 & 0.08 & 0.30 & 2.19 & 10.60 & 61.80 & 31.60 & 0.34 & 0.50 &  &  &  &  & 4.3 & 10.5 & 487 & 81.2 \\
003.7-04.6 & 0.02 & <1.3 & 2 & 1.8 & 3 & 3 & 1.0 & <42.2 &  &  &  &  &  &  &  &  & 1.39 & 1.40 &  &  &  &  & 5.6 & 5.1 & - & - \\
003.7+07.9 &  &  &  &  &  &  &  &  & 0.17 & 0.05 & 0.18 & 0.12 & 0.58 & 1.24 & 27.23 & 14.20 &  &  &  &  & 0.84 & 1.00 & 12.3 & 11.9 & 139 & 34.7 \\
003.8-04.3 &  &  &  &  &  &  &  &  & 0.14 & 0.63 & 0.14 & 0.68 & 2.87 & 7.26 & 84.97 & 57.80 & 0.60 & 0.40 &  &  &  &  & 3.4 & 6.6 & 448 & 89.6 \\
003.9-02.3 & 1 & 1.4 & 11.2 & 7.2 & 14.1 & 14.1 & 5.1 & <154 &  &  &  &  &  &  &  &  & 5.68 & 5.70 &  &  &  &  & 6.6 & 7.3 & 7 & 1.7 \\
003.9-03.1 &  &  &  &  &  &  &  &  & 0.05 & 0.11 & 0.12 & 0.12 &  &  &  &  &  &  & 0.22 & 0.19 &  &  & 7.8 & 8.1 & - & - \\
004.0-03.0 & 0.4 & 0.8 & 5.1 & 2 & 2.3 & 2.3 & 0.5 & <74.5 &  &  &  &  &  &  &  &  & 0.80 & 0.80 &  &  &  &  & 2.7 & 4.8 & 59 & 58.7 \\
004.1-03.8 &  &  &  &  &  &  &  &  &  &  &  &  &  &  &  &  & 0.25 & 0.25 &  &  &  &  & 3.0 & 3.0 & - & - \\
004.2-03.2 &  &  &  &  &  &  &  &  &  &  &  &  &  &  &  &  & 0.35 & 0.35 &  &  &  &  & 7.0 & 7.1 & - & - \\
004.2-04.3 &  &  &  &  &  &  &  &  & 0.25 & 0.28 & 0.27 & 0.19 & 2.57 & 4.54 & 65.91 & 48.90 & 1.13 & 1.13 &  &  &  &  & 4.4 & 6.0 & 21 & 20.9 \\
004.6+06.0 & 0.01 & <0.3 & 2.3 & 2.3 & 5.9 & 5.5 & 2.2 & <6.5 &  &  &  &  &  &  &  &  & 1.48 & 1.49 &  &  &  &  & 7.5 & 9.0 & 5 & 4.7 \\
004.8-05.0 & 0.2 & 0.7 & 2.6 & 0.7 & 2.6 & 2.6 & 0.8 & <44.7 &  &  &  &  &  &  &  &  & 0.78 & 0.80 &  &  &  &  & 6.0 & 11.0 & 172 & 85.8 \\
004.8+02.0 &  &  &  &  &  &  &  &  & 0.13 & 0.49 & 0.14 & 0.58 & 3.25 & 8.72 & 177 & 158 & 0.59 & 0.40 &  &  &  &  & 4.3 & 3.1 & 322 & 160.9 \\
005.0-03.9 &  &  &  &  &  &  &  &  & 0.09 & 0.24 & 0.10 & 0.38 &  &  &  &  &  &  &  &  & 0.48 & 0.31 & 7.2 & 13.0 & 376 & 376.2 \\
005.2+05.6 & 0.01 & <0.4 & 1.3 & 0.9 & 4 & 4 & 1.6 & <5.5 &  &  &  &  &  &  &  &  & 1.52 & 1.50 &  &  &  &  & 7.0 & 6.0 & 5 & 2.3 \\
005.5-04.0 & 0.005 & <1.6 & 0.7 & 0.7 & 1.3 & 1.2 & 0.5 & <81.5 &  &  &  &  &  &  &  &  &  &  & 0.46 & 0.46 &  &  & 6.1 & 6.6 & 2 & 2.0 \\
005.5+06.1 &  &  &  &  &  &  &  &  & 0.40 & 0.41 & 0.43 & 0.48 & 1.29 & 8.22 & 53.16 & 39.70 & 1.81 & 1.10 &  &  &  &  & 7.6 & 7.2 & 490 & 490.0 \\
005.8-06.1 & 0.3 & 1.6 & 7.1 & 2 & 4.4 & 4 & 1.1 & <25.6 &  &  &  &  &  &  &  &  & 0.92 & 1.00 &  &  &  &  & 2.2 & 7.4 & 449 & 49.9 \\
006.1+08.3 & 0.6 & 1.1 & 5.8 & 3.9 & 2.4 & 2.4 & 0.5 & <4.6 &  &  &  &  &  &  &  &  & 5.33 & 5.10 &  &  &  &  & 1.5 & 2.5 & 23 & 11.6 \\
006.4-04.6 & 0.003 & <2.3 & 0.9 & 0.7 & 4.2 & 1.6 & 1.9 & 1.9 &  &  &  &  &  &  &  &  &  &  & 0.52 & 0.54 &  &  & 9.3 & 8.0 & 34 & 34.1 \\
006.4+02.0 & 0.9 & 1.2 & 18.5 & 11.7 & 11.2 & 11.1 & 2.7 & <34.9 &  &  &  &  &  &  &  &  & 5.65 & 5.70 &  &  &  &  & 3.2 & 3.5 & 81 & 20.1 \\
006.8-03.4 &  &  &  &  &  &  &  &  & 0.28 & 0.73 & 0.29 & 0.56 & 2.83 & 5.14 & 86.92 & 63.80 &  &  & 1.26 & 1.08 &  &  & 4.4 & 4.6 & 95 & 95.0 \\
006.8+02.3 &  &  &  &  &  &  &  &  & 0.87 & 0.96 & 0.92 & 1.00 & 7.52 & 6.00 & 56.82 & 56.60 & 3.79 & 3.85 &  &  &  &  & 6.6 & 4.0 & 15 & 3.6 \\
007.0-06.8 & 0.2 & <0.3 & 3.5 & 3.9 & 7.9 & 6.4 & 5.4 & 5.4 &  &  &  &  &  &  &  &  & 3.75 & 3.75 &  &  &  &  & 3.7 & 4.0 & 2 & 0.5 \\
007.0+06.3 & 0.3 & 0.4 & 14.1 & 7.7 & 20.2 & 11.5 & 6.6 & 6.5 &  &  &  &  &  &  &  &  &  &  & 3.99 & 4.06 &  &  & 7.8 & 6.4 & 19 & 9.4 \\
007.5+07.4 &  &  &  &  &  &  &  &  & 0.09 & 0.22 & 0.09 & 0.33 & 2.55 & 9.22 & 56.56 & 16.80 & 0.39 & 0.35 &  &  &  &  & 9.8 & 9.0 & 273 & 45.5 \\
007.6+06.9 & 0.02 & <0.3 & 1.9 & 1.2 & 3.4 & 3.4 & 1.1 & <4.1 &  &  &  &  &  &  &  &  &  &  & 1.35 & 1.36 &  &  & 5.8 & 7.0 & 5 & 1.6 \\
007.8-03.7 &  &  &  &  &  &  &  &  & 0.15 & 0.86 & 0.15 & 0.69 & 2.06 & 9.43 & 46.89 & 23.90 &  &  &  &  & 0.73 & 0.89 & 4.5 & 8.0 & 640 & 127.9 \\
007.8-04.4 & 0.2 & <0.3 & 4.5 & 6.2 & 7.1 & 7 & 2.6 & <49 &  &  &  &  &  &  &  &  & 1.70 & 1.70 &  &  &  &  & 6.5 & 8.0 & - & - \\
008.2+06.8 & 0.7 & 0.6 & 5.7 & 8.1 & 2.9 & 2.9 & 0.7 & <4.2 &  &  &  &  &  &  &  &  & 1.30 & 1.30 &  &  &  &  & 1.7 & 1.8 & - & - \\
008.4-03.6 & 0.01 & <1.2 & 1.2 & 0.6 & 2.2 & 2.2 & 0.8 & <60.9 &  &  &  &  &  &  &  &  &  &  & 0.86 & 0.93 &  &  & 6.9 & 7.6 & 19 & 18.8 \\
008.6-02.6 &  &  &  &  &  &  &  &  & 0.17 & 0.35 & 0.18 & 0.27 & 1.04 & 7.38 & 47.10 & 29.00 &  &  & 0.80 & 1.20 &  &  & 7.7 & 8.4 & 615 & 614.6 \\
009.4-09.8 & 0.01 & <0.2 & 1.1 & 1.3 & 5.5 & 3.9 & 2.5 & 2.5 &  &  &  &  &  &  &  &  & 1.20 & 1.20 &  &  &  &  & 8.0 & 8.1 & 3 & 1.5 \\
009.8-04.6 & 0.4 & <0.3 & 4.2 & 1.1 & 1.8 & 1.8 & 0.4 & <37.6 &  &  &  &  &  &  &  &  & 1.00 & 1.10 &  &  &  &  & 26.2 & 7.0 & 318 & 79.5 \\
351.1+04.8 & 0.02 & <0.4 & 2.5 & 3.7 & 4.9 & 4.8 & 1.7 & <6.6 &  &  &  &  &  &  &  &  & 2.61 & 2.60 &  &  &  &  & 6.2 & 8.0 & 5 & 1.7 \\
351.2+05.2 & 0.3 & 0.5 & 3.8 & 1.7 & 5.3 & 5.3 & 1.9 & <6.8 &  &  &  &  &  &  &  &  & 1.19 & 1.20 &  &  &  &  & 6.2 & 6.5 & 29 & 29.4 \\
351.6-06.2 & 0.01 & <0.2 & 1 & 1.2 & 2.2 & 2 & 0.8 & <17.1 &  &  &  &  &  &  &  &  &  &  & 1.26 & 1.27 &  &  & 7.8 & 8.6 & 5 & 0.8 \\
351.9-01.9 &  &  &  &  &  &  &  &  & 0.83 & 3.34 & 0.88 & 2.38 & 20.66 & 67.10 & 531 & 518 &  &  & 3.81 & 2.90 &  &  & 3.1 & 10.0 & 353 & 88.1 \\
351.9+09.0 & 0.01 & <0.3 & 0.7 & 0.9 & 2 & 1.6 & 0.7 & <21.4 &  &  &  &  &  &  &  &  &  &  & 1.00 & 1.00 &  &  & 8.6 & 10.0 & 3 & 1.5 \\
352.0-04.6 & 0.01 & <0.3 & 1.1 & 1.1 & 2.8 & 2.6 & 1.1 & <35.4 &  &  &  &  &  &  &  &  &  &  & 0.94 & 0.96 &  &  & 7.8 & 5.4 & 11 & 1.5 \\
352.1+05.1 & 0.03 & <0.3 & 2.9 & 2.1 & 5.2 & 5.2 & 1.7 & <6.6 &  &  &  &  &  &  &  &  & 1.79 & 1.80 &  &  &  &  & 8.0 & 5.0 & 27 & 3.9 \\
352.6+03.0 &  &  &  &  &  &  &  &  & 0.49 & 1.14 & 0.53 & 1.52 & 20.55 & 50.70 & 375 & 333 & 2.23 & 1.80 &  &  &  &  & 2.7 & 3.4 & 158 & 26.3 \\
353.2-05.2 &  &  &  &  &  &  &  &  & 0.02 & 0.05 & 0.02 & 0.08 & 0.20 & 1.50 & 9.38 & 2.24 &  &  &  &  & 0.09 & 0.25 & 8.8 & 14.0 & 2419 & 302.3 \\
353.3+06.3 &  &  &  &  &  &  &  &  & 0.39 & 0.44 & 0.41 & 0.59 & 4.08 & 9.21 & 153 & 120 & 1.75 & 1.70 &  &  &  &  & 4.0 & 8.0 & 53 & 26.5 \\
353.7+06.3 & 0.1 & <0.5 & 2.1 & 3.4 & 2 & 1.2 & 0.6 & <19.8 &  &  &  &  &  &  &  &  & 0.70 & 0.70 &  &  &  &  & 6.2 & 7.8 & 20 & 5.0 \\
354.5+03.3 & 0.6 & 1.3 & 6.7 & 10 & 7.6 & 7.5 & 2.5 & <39.2 &  &  &  &  &  &  &  &  &  &  & 2.12 & 2.09 &  &  & 5.8 & 14.0 & 76 & 38.0 \\
354.9+03.5 & 0.004 & <1.4 & 0.7 & 0.8 & 3.3 & 2.7 & 1.5 & <30 &  &  &  &  &  &  &  &  &  &  & 0.98 & 0.99 &  &  &  &  & 4 & 4.1 \\
355.4-02.4 & 0.1 & <4.4 & 6.8 & 3.3 & 7.3 & 7.3 & 2.1 & <13.7 &  &  &  &  &  &  &  &  & 2.78 & 3.00 &  &  &  &  & 18.9 & 8.0 & 67 & 13.3 \\
355.9-04.2 & 0.1 & <0.5 & 7.8 & 5.3 & 8.2 & 8.2 & 2.3 & <34.5 &  &  &  &  &  &  &  &  & 3.09 & 3.10 &  &  &  &  & 3.4 & 3.5 & 4 & 1.2 \\
355.9+03.6 & 0.2 & 2.7 & 5.7 & 14.3 & 3.9 & 2.9 & 1.0 & <21 &  &  &  &  &  &  &  &  & 1.16 & 2.50 &  &  &  &  & 3.1 & 5.0 & 2777 & 694.4 \\
356.1-03.3 &  &  &  &  &  &  &  &  & 0.11 & 0.26 & 0.12 & 0.18 &  &  &  &  &  &  & 0.51 & 0.43 &  &  & 6.2 & 5.5 & 67 & 16.8 \\
356.3-06.2 &  &  &  &  &  &  &  &  & 0.10 & 0.18 & 0.10 & 0.10 &  &  &  &  &  &  &  &  &  &  & 4.5 & 9.8 & 63 & 31.3 \\
356.5-03.6 & 0.002 & <2.3 & 0.5 & <4.6 & 9 & 5 & 9.2 & 18 &  &  &  &  &  &  &  &  &  &  & 0.58 & 0.59 &  &  & 6.6 & 5.2 & 21 & 7.1 \\
356.8-05.4 &  &  &  &  &  &  &  &  & 0.04 & 0.02 & 0.04 & 0.04 &  &  &  &  &  &  & 0.17 & 0.30 &  &  & 2.8 & 7.0 & 304 & 304.2 \\
356.8+03.3 & 0.03 & <2.5 & 2.3 & 2.2 & 5.5 & 5.9 & 2.8 & <25.4 &  &  &  &  &  &  &  &  & 0.45 & 0.45 &  &  &  &  & 2.1 & 2.0 & 1 & 0.6 \\
356.9+04.4 & 0.8 & 1.2 & 13.7 & 6.2 & 8.7 & 8.7 & 2.2 & <16.4 &  &  &  &  &  &  &  &  & 2.09 & 2.10 &  &  &  &  & 1.8 & 1.6 & 29 & 4.8 \\
357.0+02.4 &  &  &  &  &  &  &  &  & 0.68 & 9.10 & 0.72 & 5.30 & 3.77 & 15.70 & 101 & 92.10 & 2.97 & 0.98 &  &  &  &  & 4.6 & 6.3 & 3963 & 792.6 \\
357.1-04.7 & 0.9 & 0.5 & 14.8 & 9.1 & 8.1 & 8.1 & 1.9 & <27.9 &  &  &  &  &  &  &  &  & 0.53 & 0.60 &  &  &  &  & 7.9 & 2.0 & - & - \\
357.1+03.6 &  &  &  &  &  &  &  &  & 1.19 & 8.70 & 1.26 & 4.80 & 11.79 & 32.50 & 453 & 420 & 5.33 & 2.80 &  &  &  &  & 5.8 & 6.5 & 894 & 447.1 \\
357.1+04.4 & 0.002 & <2.1 & 0.6 & 0.6 & 2 & 1.7 & 0.8 & <21.2 &  &  &  &  &  &  &  &  & 0.42 & 0.42 &  &  &  &  & 8.7 & 10.9 & 7 & 3.5 \\
357.2+02.0 &  &  &  &  &  &  &  &  & 0.34 & 0.27 & 0.37 & 0.53 & 5.21 & 22.70 & 168 & 120 &  &  & 1.57 & 2.05 &  &  & 5.3 & 5.6 & 181 & 90.4 \\
357.3+04.0 & 0.01 & <1.8 & 1.3 & 1.3 & 2.9 & 2.9 & 1.0 & <13.4 &  &  &  &  &  &  &  &  & 0.90 & 0.90 &  &  &  &  & 5.7 & 5.7 & 9 & 8.8 \\
357.5+03.1 &  &  &  &  &  &  &  &  & 0.12 & 0.64 & 0.13 & 0.45 & 0.19 & 0.92 & 40.18 & 34.80 &  &  & 0.54 & 0.31 &  &  & 15.2 & 6.0 & - & - \\
357.5+03.2 & 0.01 & <1.8 & 1.3 & 1.4 & 3.2 & 2.8 & 1.2 & <26.5 &  &  &  &  &  &  &  &  & 0.59 & 0.59 &  &  &  &  & 6.9 & 7.2 & 5 & 0.7 \\
357.6-03.3 &  &  &  &  &  &  &  &  & 0.17 & 0.08 & 0.18 & 0.10 & 1.23 & 5.96 & 63.86 & 42.80 &  &  & 0.78 & 0.64 &  &  & 11.4 & 10.7 & 215 & 107.3 \\
357.9-03.8 & 0.004 & <1.9 & 0.8 & 1 & 2.3 & 1.8 & 0.9 & <22.5 &  &  &  &  &  &  &  &  &  &  &  &  &  &  & 11.8 & 13.3 & 12 & 11.6 \\
357.9-05.1 & 0.004 & <0.3 & 0.3 & 0.4 & 4.3 & 3.5 & 2.7 & <27.5 &  &  &  &  &  &  &  &  & 1.47 & 1.47 &  &  &  &  & 21.6 & 29.0 & 7 & 1.0 \\
358.0+09.3 &  &  &  &  &  &  &  &  &  &  & 0.03 & 0.06 &  &  &  &  &  &  &  &  & 0.16 & 0.14 & 9.3 & 10.0 & - & - \\
358.2+03.5 &  &  &  &  &  &  &  &  & 0.37 & 0.60 & 0.39 & 0.71 & 6.85 & 14.90 & 166 & 150 & 1.66 & 1.53 &  &  &  &  & 2.1 & 3.7 & - & - \\
358.2+04.2 & 0.4 & 0.7 & 8.2 & 4 & 5 & 5 & 1.2 & <14.6 &  &  &  &  &  &  &  &  & 2.18 & 2.20 &  &  &  &  & 3.6 & 5.0 & 41 & 10.4 \\
358.5-04.2 & 0.1 & <2.6 & 2.9 & 5.7 & 2.8 & 2.7 & 0.8 & <47.3 &  &  &  &  &  &  &  &  & 4.34 & 4.26 &  &  &  &  & 3.3 & 3.0 & 22 & 7.3 \\
358.5+02.9 &  &  &  &  &  &  &  &  & 0.07 & 0.08 & 0.07 & 0.05 & 0.71 & 1.25 & 21.73 & 15.80 &  &  & 0.31 & 0.31 &  &  & 2.7 & 4.2 & 39 & 39.2 \\
358.6-05.5 &  &  &  &  &  &  &  &  & 0.05 & 0.19 & 0.05 & 0.13 & 0.23 & 0.96 & 6.69 & 2.16 &  &  &  &  & 0.24 & 0.32 & 9.3 & 20.9 & 483 & 80.5 \\
358.6+07.8 &  &  & 0.5 & <0.3 & 2.4 & 2.4 & 1.1 & <4.8 &  &  &  &  &  &  &  &  & 0.35 & 0.35 &  &  &  &  & 5.9 & 4.2 & 24 & 23.9 \\
358.7+05.2 & 0.1 & <0.7 & 6.6 & 6.2 & 6.9 & 6.7 & 1.9 & <10.9 &  &  &  &  &  &  &  &  & 1.79 & 1.80 &  &  &  &  & 3.7 & 2.2 & 13 & 12.8 \\
358.8+03.0 & 0.9 & 2.4 & 7.6 & 4.3 & 9.6 & <2.6 & 5.4 & <29.7 &  &  &  &  &  &  &  &  & 0.96 & 1.00 &  &  &  &  & 3.7 & 9.1 & 173 & 34.6 \\
358.9+03.4 & 0.4 & <2 & 6.1 & 6.2 & 6.5 & 6.5 & 2.3 & <22 &  &  &  &  &  &  &  &  & 2.59 & 2.59 &  &  &  &  & 2.7 & 2.6 & 3 & 0.7 \\
359.0-04.1 &  &  &  &  &  &  &  &  & 0.05 & 0.34 & 0.05 & 0.27 & 0.51 & 2.05 & 16.92 & 4.23 &  &  &  &  & 0.24 & 0.25 & 4.9 & 5.4 & 939 & 156.5 \\
359.1-02.9 &  &  &  &  &  &  &  &  & 0.43 & 4.92 & 0.46 & 2.51 &  &  &  &  &  &  & 2.03 & 0.60 &  &  & 12.4 & 15.5 & 5315 & 885.8 \\
359.2+04.7 & 0.1 & <1.5 & 3.8 & 4.9 & 4.2 & 4.2 & 1.2 & <9.7 &  &  &  &  &  &  &  &  & 0.40 & 0.40 &  &  &  &  & 1.5 & 1.7 & 3 & 0.9 \\
359.3-01.8 & 1.1 & 1.1 & 16.2 & 16.3 & 17.1 & 17 & 5.3 & <157 &  &  &  &  &  &  &  &  & 3.50 & 3.50 &  &  &  &  & 4.4 & 4.4 & 1 & 0.9 \\
359.6-04.8 &  &  &  &  &  &  &  &  & 0.08 & 0.04 & 0.08 & 0.10 & 1.80 & 3.47 & 69.72 & 35.80 &  &  &  &  & 0.37 & 0.37 & 8.8 & 17.7 & 63 & 31.5 \\
359.7-01.8 &  &  &  &  &  &  &  &  & 1.42 & 9.46 & 1.49 & 12.70 & 1.92 & 13.40 & 105 & 97.60 & 6.18 & 2.43 &  &  &  &  & 16.7 & 7.1 & - & - \\
359.8-07.2 & 0.6 & 4 & 5.5 & 4.5 & 2 & 1.8 & 0.4 & <13.1 &  &  &  &  &  &  &  &  &  &  & 0.79 & 0.79 &  &  & 1.2 & 8.0 & 1122 & 1122.3 \\
359.8+02.4 & 0.7 & 0.9 & 6.5 & 4.8 & 8.7 & 8.7 & 3.2 & <30.4 &  &  &  &  &  &  &  &  &  &  & 1.23 & 1.23 &  &  & 5.4 & 6.0 & 9 & 8.7 \\
359.8+03.7 & 0.03 & <3.2 & 1.7 & <4.4 & 1.8 & 1.8 & 0.5 & <11.8 &  &  &  &  &  &  &  &  & 1.80 & 1.80 &  &  &  &  & 3.1 & 3.0 & 5 & 5.3 \\
359.8+05.2 &  &  &  &  &  &  &  &  & 0.01 & 0.02 & 0.01 & 0.02 &  &  &  &  &  &  &  &  & 0.05 & 0.05 & 12.9 & 19.0 & 40 & 19.8 \\
359.8+05.6 & 0.1 & <0.4 & 4.5 & 4.7 & 4.4 & 3.9 & 1.2 & <6.2 &  &  &  &  &  &  &  &  & 1.27 & 1.29 &  &  &  &  & 5.2 & 4.4 & 4 & 4.0 \\
359.8+06.9 &  &  &  &  &  &  &  &  & 0.10 & 0.19 & 0.11 & 0.30 & 0.93 & 2.47 & 38.22 & 9.70 & 0.45 & 0.50 &  &  &  &  & 7.6 & 10.0 & 215 & 26.9 \\
359.9-04.5 & 0.8 & 1.1 & 11.1 & 7.1 & 6.9 & 6.9 & 1.7 & <51.3 &  &  &  &  &  &  &  &  & 2.90 & 2.90 &  &  &  &  & 2.6 & 3.3 & 25 & 5.1 \\
\end{longtable}
\normalsize
\twocolumn
 
\onecolumn
\small
\begin{longtable}{crrrrrccccccc}
\caption{Outputs of the {\scshape cloudy} models for 124 PNe. The table includes key parameters such as the PN G designation, effective temperature, luminosity, inner radius, hydrogen density, and dust-to-gas ratio. Elemental abundances relative to hydrogen are also provided, with a "-" marker indicating unphysical abundances.}
\label{T:cloudy_out}\\
\hline
    PN G & T$_{eff}$ & L & R$_{in}$ & n$_{H}$ & $\rho_{d}$/$\rho_{g}$ & He/H & N/H & O/H & Ne/H & S/H & Cl/H & Ar/H \\
    &(K) & (\(L_\odot\)) & (pc) & (cm$^{-3}$) & &  &  & &  &  & &  \\
\hline
\endfirsthead

\multicolumn{13}{c}%
{{\bfseries \tablename\ \thetable{} -- continued}} \\
\hline
    PN G & T$_{eff}$ & L & R$_{in}$ & n$_{H}$ & $\rho_{d}$/$\rho_{g}$ & He/H & N/H & O/H & Ne/H & S/H & Cl/H & Ar/H \\
    &(K) & (\(L_\odot\)) & (pc) & (cm$^{-3}$) & &  &  & &  &  & &  \\
\hline
\endhead

\hline
\endfoot

\hline \hline
\endlastfoot

000.1-02.3 & 156124 & 3843 & 0.100 & 33467 & 0.85 & 0.49 & -4.16 & -2.74 & -4.13 & -5.06 & -6.73 & -5.50 \\
000.1+02.6 & 124862 & 3405 & 0.112 & 5861 & 0.82 & -0.13 & -4.42 & -3.65 & -4.13 & -4.21 & -6.91 & -5.90 \\
000.1+04.3 & 108657 & 2883 & 0.103 & 3881 & 0.20 & -0.88 & -4.70 & -3.31 & -3.82 & -5.27 & -6.93 & -5.72 \\
000.2-01.9 & 100597 & 2629 & 0.267 & 7457 & 1.21 & -0.86 & -4.61 & -2.85 & -4.16 & -5.50 & -6.94 & -5.76 \\
000.2-04.6 & 95816 & 10113 & 0.110 & 63095 & 0.95 & -0.82 & -3.62 & -3.23 & -4.01 & -4.08 & -7.14 & -5.68 \\
000.3-04.6 & 83648 & 2524 & 0.149 & 9282 & 1.43 & -0.36 & -3.97 & -3.18 & -3.99 & -4.92 & -6.21 & -5.74 \\
000.3+06.9 & 90455 & 3306 & 0.309 & 3021 & 0.89 & -0.85 & -4.15 & -3.65 & -4.30 & -5.24 & - & -5.99 \\
000.4-01.9 & 131608 & 3511 & 0.094 & 2683 & 1.47 & -1.10 & -3.81 & -4.04 & -4.19 & -4.72 & -6.56 & -5.91 \\
000.4-02.9 & 138146 & 2708 & 0.144 & 8067 & 2.01 & -0.17 & -5.19 & -3.23 & -4.56 & -5.76 & -7.28 & -5.23 \\
000.7-02.7 & 121686 & 2179 & 0.026 & 63095 & 1.44 & 0.34 & -4.90 & -3.23 & -4.18 & -5.91 & -7.60 & -6.16 \\
000.7-07.4 & 65901 & 5621 & 0.099 & 63090 & 1.16 & -0.92 & -3.76 & -2.77 & -3.39 & -4.16 & -6.87 & -6.08 \\
000.7+03.2 & 100537 & 3849 & 0.131 & 3143 & 0.89 & -0.79 & -2.72 & -2.65 & -2.82 & -4.29 & -6.15 & -4.94 \\
000.9-02.0 & 76803 & 3109 & 0.066 & 63047 & 1.39 & -0.62 & -4.95 & -3.00 & -4.01 & -5.59 & -7.07 & -5.85 \\
000.9-04.8 & 79163 & 10108 & 0.079 & 6763 & 0.49 & -0.35 & -3.86 & -3.59 & -4.08 & -4.48 & -6.96 & -5.62 \\
001.1-01.6 & 95752 & 2942 & 0.254 & 3358 & 0.88 & -0.64 & -3.61 & -2.96 & -3.64 & -4.76 & -6.85 & -5.47 \\
001.2-03.0 & 143843 & 2589 & 0.035 & 4877 & 0.86 & -1.95 & -3.98 & -4.36 & - & -5.35 & - & -6.21 \\
001.2+02.1 & 86960 & 2783 & 0.043 & 20819 & 0.27 & -0.92 & -4.18 & -3.70 & -4.38 & -4.47 & -7.10 & -5.52 \\
001.3-01.2 & 195721 & 2893 & 0.063 & 8511 & 1.40 & -0.23 & -3.64 & -5.00 & - & -6.66 & -6.66 & -6.51 \\
001.4+05.3 & 96832 & 2615 & 0.088 & 27863 & 1.43 & -0.68 & -5.14 & -3.57 & -6.04 & -7.14 & -7.45 & -7.19 \\
001.6-01.3 & 251174 & 622 & 0.039 & 4713 & 1.22 & -1.48 & -4.72 & -2.90 & -4.04 & -3.76 & -6.43 & -5.46 \\
001.7-04.4 & 153120 & 2967 & 0.096 & 3793 & 1.31 & -0.66 & -3.65 & -5.47 & - & -6.13 & -6.66 & -6.10 \\
001.7+05.7 & 104889 & 10067 & 0.120 & 13512 & 0.16 & -1.20 & -5.87 & -3.55 & -4.20 & -7.49 & -7.66 & -7.28 \\
002.0-06.2 & 79384 & 3059 & 0.082 & 28758 & 0.99 & -1.08 & -5.74 & -3.46 & -4.45 & -6.75 & -6.71 & -6.68 \\
002.1-02.2 & 90199 & 3533 & 0.055 & 32371 & 0.45 & -0.65 & -4.55 & -3.39 & -4.02 & -5.84 & -7.12 & -6.44 \\
002.1-04.2 & 89418 & 7460 & 0.030 & 63033 & 0.27 & -1.61 & -3.42 & -4.04 & -4.81 & -7.46 & -7.55 & -7.56 \\
002.2-09.4 & 99355 & 2128 & 0.003 & 1494 & 0.45 & -1.12 & -3.55 & -3.65 & -4.17 & -5.10 & -6.70 & -5.62 \\
002.3+02.2 & 90072 & 7750 & 0.118 & 63096 & 1.33 & -0.67 & -4.25 & -3.30 & -3.91 & -5.45 & -7.45 & -5.71 \\
002.5-01.7 & 101622 & 10021 & 0.100 & 55734 & 0.93 & -0.89 & -3.88 & -3.58 & -4.28 & -4.36 & -6.71 & -5.87 \\
002.6+02.1 & 114103 & 3350 & 0.067 & 15212 & 1.03 & -0.92 & -3.37 & -3.33 & -3.66 & -4.28 & -6.96 & -5.45 \\
002.7-04.8 & 92272 & 1764 & 0.202 & 2050 & 1.49 & -0.39 & -4.37 & -3.62 & -4.23 & -4.48 & -6.92 & -5.39 \\
002.8+01.7 & 69505 & 3706 & 0.121 & 41643 & 0.38 & -0.66 & -4.20 & -3.66 & - & -5.16 & -6.84 & -5.65 \\
002.8+01.8 & 78026 & 3525 & 0.119 & 63095 & 1.27 & -0.44 & -3.72 & -2.25 & -3.62 & -4.17 & -8.03 & -5.51 \\
002.9-03.9 & 140732 & 3274 & 0.093 & 19054 & 1.00 & -0.42 & -4.30 & -3.49 & -4.22 & -5.21 & -7.09 & -5.87 \\
003.2-06.2 & 95337 & 2052 & 0.045 & 29554 & 1.38 & -0.34 & -4.00 & -3.79 & -4.48 & -5.01 & -6.40 & -5.85 \\
003.6-02.3 & 91516 & 3924 & 0.067 & 63050 & 1.37 & -0.81 & -4.23 & -3.61 & -4.29 & -5.17 & -7.04 & -5.41 \\
003.7-04.6 & 107814 & 2999 & 0.080 & 11431 & 0.91 & -1.03 & -5.76 & -3.53 & -5.96 & -7.11 & -7.05 & -7.01 \\
003.7+07.9 & 118936 & 5378 & 0.082 & 5921 & 0.17 & -0.90 & -2.48 & -3.16 & -3.38 & -4.14 & - & -6.44 \\
003.8-04.3 & 121731 & 4492 & 0.061 & 44846 & 1.18 & -0.75 & -4.40 & -3.60 & -4.31 & -5.57 & -6.92 & -5.93 \\
003.9-02.3 & 73296 & 2959 & 0.003 & 2332 & 0.00 & -1.03 & -3.90 & -3.77 & -4.28 & -4.97 & -6.75 & -5.73 \\
003.9-03.1 & 87910 & 2657 & 0.021 & 565 & 1.55 & -0.93 & -4.38 & -3.02 & -4.20 & -5.53 & -7.18 & -5.97 \\
004.0-03.0 & 48622 & 8480 & 0.063 & 22678 & 0.79 & -0.90 & -5.34 & -4.26 & -4.45 & -7.11 & -8.39 & -7.06 \\
004.1-03.8 & 98137 & 3052 & 0.032 & 2186 & 1.88 & -1.00 & -4.52 & -3.52 & -4.31 & -3.71 & -7.33 & -5.62 \\
004.2-03.2 & 152587 & 2975 & 0.218 & 635 & 1.54 & -0.96 & -5.19 & -3.44 & -4.33 & -6.26 & -7.10 & -6.19 \\
004.2-04.3 & 63574 & 3634 & 0.064 & 30865 & 0.54 & -0.79 & -4.11 & -3.37 & -4.03 & -5.35 & -7.23 & -6.88 \\
004.6+06.0 & 185157 & 3686 & 0.114 & 4169 & 1.06 & -0.75 & -4.55 & -4.30 & -5.46 & -6.63 & -7.03 & -6.54 \\
004.8-05.0 & 91819 & 3141 & 0.003 & 879 & -0.05 & -1.26 & -3.96 & -3.48 & -3.91 & -5.11 & -6.80 & -5.46 \\
004.8+02.0 & 98747 & 2847 & 0.092 & 62635 & 1.77 & -1.23 & -4.71 & -3.32 & - & -6.21 & -6.61 & -6.28 \\
005.0-03.9 & 98687 & 2520 & 0.057 & 501 & 1.72 & -0.57 & -5.09 & -3.99 & -4.44 & -6.09 & -8.70 & -6.48 \\
005.2+05.6 & 108036 & 2195 & 0.107 & 3340 & 0.73 & -1.04 & -4.29 & -3.91 & -4.37 & -4.62 & -6.54 & -5.27 \\
005.5-04.0 & 137678 & 4304 & 0.114 & 5635 & 1.11 & -0.63 & -5.77 & -3.81 & - & -7.17 & -7.21 & -6.99 \\
005.5+06.1 & 99659 & 2630 & 0.133 & 55775 & 0.96 & -0.77 & -4.40 & -3.23 & -4.18 & -5.46 & -7.05 & -6.19 \\
005.8-06.1 & 96487 & 3368 & 0.038 & 63091 & 1.70 & -0.77 & -3.55 & -3.03 & -3.93 & -3.92 & -6.62 & -5.48 \\
006.1+08.3 & 85232 & 2941 & 0.026 & 51218 & -0.01 & -1.05 & -4.49 & -3.51 & -4.21 & -5.57 & -7.17 & -6.07 \\
006.4-04.6 & 136785 & 2044 & 0.141 & 2447 & 1.48 & -0.62 & -4.11 & -3.64 & - & -5.02 & -6.45 & -5.40 \\
006.4+02.0 & 66518 & 2910 & 0.037 & 17934 & 0.72 & -0.68 & -3.62 & -3.76 & -4.40 & -5.84 & -6.77 & -5.82 \\
006.8-03.4 & 79588 & 3143 & 0.064 & 41391 & 0.84 & -0.82 & -4.90 & -3.43 & -4.19 & -6.12 & -7.64 & -6.19 \\
006.8+02.3 & 132715 & 8967 & 0.032 & 63092 & -0.52 & -1.04 & -3.13 & -3.32 & -4.01 & -6.55 & -6.95 & -6.57 \\
007.0-06.8 & 208606 & 1860 & 0.004 & 2477 & -0.20 & -0.07 & -4.05 & -3.67 & -4.20 & -4.23 & -6.48 & -4.38 \\
007.0+06.3 & 120008 & 2751 & 0.028 & 1793 & 0.48 & -1.36 & -3.82 & -3.75 & -4.47 & -4.56 & -6.66 & -5.37 \\
007.5+07.4 & 94681 & 10116 & 0.082 & 63071 & 1.08 & -1.01 & -3.76 & -3.61 & -4.19 & -4.18 & -7.05 & -5.85 \\
007.6+06.9 & 110660 & 2773 & 0.090 & 9241 & 0.98 & -0.91 & -4.41 & -3.84 & -4.29 & -5.52 & -6.85 & -5.52 \\
007.8-03.7 & 110633 & 6706 & 0.066 & 63089 & 0.86 & -0.88 & -3.88 & -3.72 & -4.27 & -5.26 & -3.99 & -6.05 \\
007.8-04.4 & 71756 & 3047 & 0.014 & 1109 & 0.31 & -1.50 & -3.85 & -4.05 & - & -5.69 & - & -7.09 \\
008.2+06.8 & 56509 & 4933 & 0.018 & 6561 & 0.16 & -2.17 & -4.01 & -4.73 & - & -6.65 & -8.14 & -6.98 \\
008.4-03.6 & 136311 & 3926 & 0.094 & 4622 & 0.79 & -0.86 & -3.41 & -4.96 & -3.11 & -5.07 & -7.20 & -5.48 \\
008.6-02.6 & 73625 & 3341 & 0.111 & 44386 & 1.20 & -0.62 & - & -2.73 & -4.00 & -6.04 & -7.41 & -6.41 \\
009.4-09.8 & 117763 & 2324 & 0.116 & 915 & 0.56 & -1.22 & -4.00 & -3.83 & -4.30 & -5.05 & -6.87 & -5.63 \\
009.8-04.6 & 101054 & 2433 & 0.022 & 15547 & 0.49 & -1.01 & -2.88 & -3.40 & -4.04 & -4.73 & -6.54 & -5.76 \\
351.1+04.8 & 63057 & 2454 & 0.105 & 7191 & 0.75 & -1.02 & -4.21 & -3.26 & -3.96 & -4.49 & -6.73 & -5.32 \\
351.2+05.2 & 82059 & 2268 & 0.003 & 991 & 0.33 & -0.91 & -2.91 & -5.21 & -4.15 & -5.44 & -6.79 & -5.77 \\
351.6-06.2 & 108062 & 3458 & 0.115 & 2887 & 0.57 & -0.69 & -3.12 & -3.72 & -4.26 & -4.60 & -7.10 & -5.63 \\
351.9-01.9 & 60651 & 3230 & 0.052 & 7910 & 0.48 & -0.29 & -3.93 & -3.48 & -3.89 & -5.83 & -6.93 & -5.85 \\
351.9+09.0 & 144140 & 4511 & 0.140 & 5365 & 0.92 & -0.81 & -5.58 & -3.66 & -4.21 & -6.46 & -7.28 & -6.52 \\
352.0-04.6 & 114613 & 2826 & 0.106 & 3970 & 0.83 & -1.25 & -2.88 & -3.29 & -3.71 & -4.55 & -7.01 & -5.39 \\
352.1+05.1 & 94879 & 1774 & 0.070 & 6961 & 0.89 & -1.08 & -3.63 & -3.69 & -4.08 & -4.72 & -6.98 & -5.76 \\
352.6+03.0 & 58111 & 2651 & 0.042 & 18915 & 0.68 & -0.40 & -3.34 & -3.41 & -3.76 & -4.54 & -6.45 & -5.55 \\
353.2-05.2 & 82022 & 3943 & 0.109 & 42516 & 1.41 & -0.66 & -4.15 & -3.33 & -4.16 & -4.68 & -7.66 & -6.15 \\
353.3+06.3 & 73195 & 3037 & 0.071 & 37755 & 0.97 & -0.89 & -4.77 & -3.34 & -4.32 & -5.72 & -7.13 & -6.00 \\
353.7+06.3 & 71023 & 3916 & 0.088 & 63087 & 1.47 & -2.08 & -4.59 & -3.52 & -4.80 & -5.98 & -7.02 & -6.05 \\
354.5+03.3 & 97016 & 2762 & 0.003 & 1532 & 0.32 & -1.06 & -3.52 & -3.29 & -3.74 & -5.18 & -6.93 & -5.59 \\
354.9+03.5 & 128336 & 2948 & 0.250 & 5550 & 1.52 & -0.34 & -4.54 & -3.47 & -6.04 & -7.39 & -7.90 & -7.25 \\
355.4-02.4 & 95499 & 2450 & 0.056 & 17738 & 1.02 & -1.18 & -3.58 & -3.73 & -4.26 & -4.78 & -6.71 & -5.81 \\
355.9-04.2 & 130114 & 2920 & 0.049 & 8744 & 0.92 & -0.44 & -3.28 & -4.55 & -3.95 & -4.57 & -6.20 & -5.09 \\
355.9+03.6 & 84387 & 3044 & 0.045 & 63092 & 1.47 & -1.03 & -4.93 & -3.89 & -4.41 & -5.49 & -7.50 & -6.33 \\
356.1-03.3 & 174217 & 2460 & 0.114 & 501 & 0.80 & -1.01 & -3.57 & -3.66 & -4.14 & -5.11 & -6.10 & -5.71 \\
356.3-06.2 & 117206 & 3023 & 0.056 & 1617 & 1.85 & -0.91 & -3.68 & -3.75 & -4.18 & -4.86 & -6.97 & -5.76 \\
356.5-03.6 & 251187 & 1558 & 0.124 & 1045 & 1.33 & -0.98 & -4.45 & -3.35 & -4.20 & -3.82 & -6.72 & -5.74 \\
356.8-05.4 & 108239 & 2624 & 0.034 & 3356 & 2.17 & -0.95 & -3.84 & -3.42 & -4.16 & -5.19 & -6.88 & -5.54 \\
356.8+03.3 & 71241 & 3338 & 0.081 & 28580 & 1.62 & -1.30 & -4.24 & -4.57 & - & -5.07 & -6.40 & -5.84 \\
356.9+04.4 & 88270 & 3022 & 0.024 & 10939 & 0.92 & -0.60 & -3.60 & -3.44 & -3.96 & -5.16 & -6.99 & -5.69 \\
357.0+02.4 & 127503 & 5050 & 0.105 & 29504 & 1.06 & 0.17 & -4.33 & -3.24 & -4.24 & -5.35 & -6.80 & -5.84 \\
357.1-04.7 & 83738 & 3122 & 0.035 & 21197 & 1.59 & -1.09 & -3.07 & -3.37 & - & -4.94 & -5.78 & - \\
357.1+03.6 & 84334 & 3149 & 0.080 & 51577 & 1.18 & -0.62 & -4.86 & -3.20 & -4.51 & -4.88 & -7.13 & -5.27 \\
357.1+04.4 & 101126 & 2697 & 0.121 & 660 & 0.48 & -0.99 & -4.63 & -2.90 & -3.25 & -5.42 & -7.06 & -6.21 \\
357.2+02.0 & 96574 & 3522 & 0.072 & 63093 & 1.29 & -0.69 & -4.66 & -2.84 & -3.89 & -5.43 & -7.08 & -5.72 \\
357.3+04.0 & 149404 & 2448 & 0.062 & 1029 & 0.28 & -0.93 & -5.65 & -3.41 & -4.06 & -7.56 & -7.95 & -7.56 \\
357.5+03.1 & 102716 & 2612 & 0.276 & 4017 & 1.74 & -12.03 & -4.87 & -4.98 & - & -6.14 & -7.10 & - \\
357.5+03.2 & 113360 & 2671 & 0.097 & 2457 & 0.99 & -0.75 & -2.95 & -3.50 & -3.88 & -4.54 & -7.33 & -5.62 \\
357.6-03.3 & 99100 & 2705 & 0.148 & 63088 & 1.48 & -0.60 & -4.49 & -2.68 & -3.66 & -5.65 & - & -5.97 \\
357.9-03.8 & 112041 & 5727 & 0.192 & 5207 & 1.64 & -0.38 & -6.20 & -3.91 & -4.26 & -7.13 & -7.87 & -6.98 \\
357.9-05.1 & 104758 & 3300 & 0.324 & 736 & 0.73 & -1.17 & -3.23 & -3.54 & -3.91 & -4.78 & -7.07 & -5.74 \\
358.0+09.3 & 69905 & 2047 & 0.168 & 901 & 2.37 & -0.98 & -4.60 & -3.37 & -4.32 & -6.08 & -7.43 & -5.96 \\
358.2+03.5 & 88114 & 2942 & 0.045 & 40375 & 0.74 & -0.89 & -4.12 & -3.33 & -4.30 & -5.57 & -7.15 & -6.12 \\
358.2+04.2 & 46005 & 3320 & 0.053 & 9174 & 0.51 & -0.74 & -3.86 & -3.54 & -3.90 & -5.04 & -6.45 & -5.27 \\
358.5-04.2 & 86464 & 3004 & 0.059 & 11776 & 0.10 & -1.25 & -4.54 & -3.94 & -4.79 & -5.39 & -7.33 & -6.04 \\
358.5+02.9 & 67096 & 3600 & 0.071 & 34684 & 0.83 & -0.68 & -5.23 & -3.17 & -3.90 & -6.39 & -7.49 & -6.63 \\
358.6-05.5 & 92683 & 10113 & 0.120 & 49641 & 0.60 & -0.95 & -4.26 & -3.63 & -4.36 & -6.34 & -7.30 & -6.20 \\
358.6+07.8 & 138628 & 2133 & 0.110 & 957 & 0.76 & -1.19 & -4.11 & -3.55 & -4.24 & -4.90 & -6.44 & -5.65 \\
358.7+05.2 & 61041 & 2552 & 0.058 & 8434 & 0.79 & -1.38 & -3.50 & -4.04 & - & -6.44 & -6.35 & -6.58 \\
358.8+03.0 & 114820 & 2176 & 0.003 & 1150 & 1.11 & -0.03 & -4.13 & -3.73 & -4.27 & -4.55 & -6.59 & -5.71 \\
358.9+03.4 & 75285 & 3104 & 0.019 & 4487 & 0.03 & -0.81 & -3.14 & -2.95 & -3.51 & -4.07 & -5.77 & -5.06 \\
359.0-04.1 & 72249 & 4821 & 0.094 & 63079 & 1.18 & -0.83 & -3.63 & -2.76 & -3.51 & -4.05 & -7.66 & -5.54 \\
359.1-02.9 & 115675 & 2688 & 0.057 & 501 & 0.90 & -0.89 & -3.82 & -4.04 & -4.44 & -5.45 & -6.78 & -5.46 \\
359.2+04.7 & 65979 & 2786 & 0.047 & 9935 & 1.56 & -0.43 & -4.34 & -4.11 & - & -6.17 & -6.55 & -5.97 \\
359.3-01.8 & 181177 & 4677 & 0.007 & 3866 & 0.31 & -3.07 & -4.22 & -4.28 & - & -6.90 & -7.05 & -6.85 \\
359.6-04.8 & 95437 & 7285 & 0.126 & 28880 & 1.83 & 0.01 & -5.24 & -3.42 & -4.37 & -6.22 & -7.27 & -6.35 \\
359.7-01.8 & 143784 & 3347 & 0.207 & 3932 & 0.90 & -0.39 & -5.45 & -3.27 & -5.03 & -6.12 & -7.08 & -6.08 \\
359.8-07.2 & 82715 & 4917 & 0.027 & 54052 & 1.06 & -0.95 & -4.68 & -3.72 & -4.35 & -5.97 & -7.05 & -6.29 \\
359.8+02.4 & 88181 & 2865 & 0.003 & 1389 & 0.62 & -1.63 & -3.79 & -5.14 & - & -4.74 & -6.42 & -5.61 \\
359.8+03.7 & 113789 & 2811 & 0.059 & 17474 & 0.57 & -1.04 & -4.56 & -4.04 & - & -7.30 & -7.39 & -7.34 \\
359.8+05.2 & 108396 & 3602 & 0.233 & 501 & 2.74 & -1.83 & -3.95 & -3.91 & - & -5.54 & -6.05 & -6.19 \\
359.8+05.6 & 172922 & 4688 & 0.064 & 11291 & 1.33 & -0.40 & -3.63 & -3.65 & - & -7.60 & -7.60 & -7.39 \\
359.8+06.9 & 67044 & 6152 & 0.114 & 40771 & 1.10 & -1.02 & -3.81 & -2.82 & -3.52 & -4.48 & -7.24 & -5.67 \\
359.9-04.5 & 89746 & 2666 & 0.016 & 5297 & 0.47 & -0.44 & -3.45 & -3.64 & -3.24 & -4.62 & -6.39 & -5.57    \\                     
\hline
\end{longtable}
\normalsize
\twocolumn

\begin{table*}
\footnotesize
\centering
\caption{The final and initial progenitor masses of the 124 PNe are derived from the evolutionary tracks presented by \protect\cite{2016A&A...588A..25M} and linear interpolation tracks (see Fig. \ref{F:cloudy_hr}). The mass losses of the PNe are also included. with the last column representing the evolutionary ages following the onset of the post-AGB phase.}
\begin{tabular}{lcccr|lcccr}
\hline\hline
PN G & M$_i$ & M$_f$ & Mass Loss & $\tau$ & PN G & M$_i$ & M$_f$ & Mass Loss & $\tau$ \\
& ($M_\odot$) &($M_\odot$) & ($M_\odot$) & (yr) & & ($M_\odot$) & ($M_\odot$) &  ($M_\odot$) & (yr) \\
\hline
000.1-02.3 & 2.3  & 0.603 & 1.697 & 1454  & 007.8-04.4 & 1    & 0.528 & 0.472 & 13813 \\
000.1+02.6 & 1.3  & 0.564 & 0.736 & 4491  & 008.2+06.8 & 1.25 & 0.561 & 0.688 & 1996  \\
000.1+04.3 & 1.2  & 0.555 & 0.645 & 7661  & 008.4-03.6 & 1.5  & 0.576 & 0.924 & 3558  \\
000.2-01.9 & 1.1  & 0.541 & 0.559 & 12867 & 008.6-02.6 & 1.1  & 0.541 & 0.559 & 9494  \\
000.2-04.6 & 3    & 0.657 & 2.343 & 153   & 009.4-09.8 & 1.1  & 0.541 & 0.559 & 14178 \\
000.3-04.6 & 1    & 0.528 & 0.472 & 16399 & 009.8-04.6 & 1.1  & 0.541 & 0.559 & 13384 \\
000.3+06.9 & 1.1  & 0.541 & 0.559 & 10993 & 351.1+04.8 & 1    & 0.528 & 0.472 & 13445 \\
000.4-01.9 & 1.4  & 0.570 & 0.830 & 4119  & 351.2+05.2 & 1    & 0.528 & 0.472 & 17087 \\
000.4-02.9 & 1.3  & 0.564 & 0.736 & 5062  & 351.6-06.2 & 1.2  & 0.555 & 0.645 & 6914  \\
000.7-02.7 & 1.1  & 0.541 & 0.559 & 14592 & 351.9-01.9 & 1    & 0.528 & 0.472 & 11970 \\
000.7-07.4 & 1.5  & 0.576 & 0.924 & 1734  & 351.9+09.0 & 1.8  & 0.579 & 1.221 & 2295  \\
000.7+03.2 & 1.25 & 0.561 & 0.688 & 3905  & 352.0-04.6 & 1.2  & 0.555 & 0.645 & 7854  \\
000.9-02.0 & 1.1  & 0.541 & 0.559 & 10055 & 352.1+05.1 & 1    & 0.528 & 0.472 & 20531 \\
000.9-04.8 & 2.9  & 0.650 & 2.250 & 186   & 352.6+03.0 & 1    & 0.528 & 0.472 & 12063 \\
001.1-01.6 & 1.1  & 0.541 & 0.559 & 11915 & 353.2-05.2 & 1.2  & 0.555 & 0.645 & 5400  \\
001.2-03.0 & 1.4  & 0.570 & 0.830 & 4702  & 353.3+06.3 & 1    & 0.528 & 0.472 & 14036 \\
001.2+02.1 & 1.1  & 0.541 & 0.559 & 11613 & 353.7+06.3 & 1.2  & 0.555 & 0.645 & 4752  \\
001.3-01.2 & 2.4  & 0.611 & 1.789 & 1363  & 354.5+03.3 & 1.1  & 0.541 & 0.559 & 12356 \\
001.4+05.3 & 1.1  & 0.541 & 0.559 & 12704 & 354.9+03.5 & 1.25 & 0.561 & 0.689 & 5041  \\
001.6-01.3 & 4.2  & 0.868 & 3.332 & 179   & 355.4-02.4 & 1    & 0.528 & 0.472 & 17602 \\
001.7-04.4 & 1.6  & 0.577 & 1.023 & 3644  & 355.9-04.2 & 1.25 & 0.561 & 0.689 & 5075  \\
001.7+05.7 & 3    & 0.657 & 2.343 & 162   & 355.9+03.6 & 1.1  & 0.541 & 0.559 & 10900 \\
002.0-06.2 & 1.1  & 0.541 & 0.559 & 10395 & 356.1-03.3 & 1.9  & 0.579 & 1.321 & 2457  \\
002.1-02.2 & 1.2  & 0.555 & 0.645 & 6118  & 356.3-06.2 & 1.2  & 0.555 & 0.645 & 7634  \\
002.1-04.2 & 2.3  & 0.603 & 1.697 & 610   & 356.5-03.6 & 3.5  & 0.745 & 2.755 & 244   \\
002.2-09.4 & 1    & 0.528 & 0.472 & 19004 & 356.8-05.4 & 1.1  & 0.541 & 0.559 & 13188 \\
002.3+02.2 & 2.4  & 0.611 & 1.789 & 539   & 356.8+03.3 & 1.1  & 0.541 & 0.558 & 9238  \\
002.5-01.7 & 3    & 0.657 & 2.343 & 159   & 356.9+04.4 & 1.1  & 0.541 & 0.559 & 11256 \\
002.6+02.1 & 1.2  & 0.555 & 0.645 & 7192  & 357.0+02.4 & 1.8  & 0.579 & 1.221 & 2035  \\
002.7-04.8 & 1    & 0.528 & 0.472 & 20517 & 357.1-04.7 & 1.1  & 0.541 & 0.559 & 10730 \\
002.8+01.7 & 1.1  & 0.541 & 0.558 & 8781  & 357.1+03.6 & 1.1  & 0.541 & 0.559 & 10741 \\
002.8+01.8 & 1.1  & 0.541 & 0.559 & 9731  & 357.1+04.4 & 1.1  & 0.541 & 0.559 & 12727 \\
002.9-03.9 & 1.4  & 0.570 & 0.830 & 4307  & 357.2+02.0 & 1.2  & 0.555 & 0.645 & 6437  \\
003.2-06.2 & 1    & 0.528 & 0.472 & 19091 & 357.3+04.0 & 1.5  & 0.576 & 0.924 & 4289  \\
003.6-02.3 & 1.2  & 0.555 & 0.645 & 5899  & 357.5+03.1 & 1.1  & 0.541 & 0.559 & 13001 \\
003.7-04.6 & 1.2  & 0.555 & 0.645 & 7462  & 357.5+03.2 & 1.2  & 0.555 & 0.645 & 8063  \\
003.7+07.9 & 1.8  & 0.579 & 1.221 & 1884  & 357.6-03.3 & 1.1  & 0.541 & 0.559 & 12610 \\
003.8-04.3 & 1.5  & 0.576 & 0.924 & 3180  & 357.9-03.8 & 1.9  & 0.579 & 1.320 & 1431  \\
003.9-02.3 & 1    & 0.528 & 0.472 & 14154 & 357.9-05.1 & 1.2  & 0.555 & 0.645 & 6981  \\
003.9-03.1 & 1    & 0.528 & 0.472 & 16493 & 358.0+09.3 & 1    & 0.528 & 0.472 & 16235 \\
004.0-03.0 & 2.4  & 0.611 & 1.789 & 297   & 358.2+03.5 & 1.1  & 0.541 & 0.559 & 11381 \\
004.1-03.8 & 1.1  & 0.541 & 0.559 & 11858 & 358.2+04.2 & 1    & 0.528 & 0.472 & 9280  \\
004.2-03.2 & 1.6  & 0.577 & 1.023 & 3640  & 358.5-04.2 & 1.1  & 0.541 & 0.559 & 11135 \\
004.2-04.3 & 1.1  & 0.541 & 0.558 & 8180  & 358.5+02.9 & 1.1  & 0.541 & 0.558 & 8610  \\
004.6+06.0 & 2.2  & 0.596 & 1.604 & 1559  & 358.6-05.5 & 3    & 0.657 & 2.343 & 149   \\
004.8-05.0 & 1.1  & 0.541 & 0.559 & 11324 & 358.6+07.8 & 1.25 & 0.561 & 0.689 & 5676  \\
004.8+02.0 & 1.1  & 0.541 & 0.559 & 12266 & 358.7+05.2 & 1    & 0.528 & 0.472 & 12832 \\
005.0-03.9 & 1.1  & 0.541 & 0.559 & 13055 & 358.8+03.0 & 1.1  & 0.541 & 0.559 & 14482 \\
005.2+05.6 & 1.1  & 0.541 & 0.559 & 14321 & 358.9+03.4 & 1.1  & 0.541 & 0.559 & 9906  \\
005.5-04.0 & 1.7  & 0.578 & 1.122 & 2686  & 359.0-04.1 & 1.25 & 0.561 & 0.688 & 2615  \\
005.5+06.1 & 1.1  & 0.541 & 0.559 & 12823 & 359.1-02.9 & 1.2  & 0.555 & 0.645 & 8069  \\
005.8-06.1 & 1.2  & 0.555 & 0.645 & 6586  & 359.2+04.7 & 1    & 0.528 & 0.472 & 13356 \\
006.1+08.3 & 1.1  & 0.541 & 0.559 & 11137 & 359.3-01.8 & 2.3  & 0.603 & 1.697 & 1311  \\
006.4-04.6 & 1.25 & 0.561 & 0.689 & 5731  & 359.6-04.8 & 2.2  & 0.596 & 1.604 & 739   \\
006.4+02.0 & 1    & 0.528 & 0.472 & 13246 & 359.7-01.8 & 1.5  & 0.576 & 0.924 & 3862  \\
006.8-03.4 & 1.1  & 0.541 & 0.559 & 10306 & 359.8-07.2 & 1.4  & 0.570 & 0.830 & 2523  \\
006.8+02.3 & 2.8  & 0.642 & 2.158 & 369   & 359.8+02.4 & 1.1  & 0.541 & 0.559 & 11546 \\
007.0-06.8 & 2.9  & 0.650 & 2.250 & 589   & 359.8+03.7 & 1.2  & 0.555 & 0.645 & 7861  \\
007.0+06.3 & 1.2  & 0.555 & 0.645 & 8037  & 359.8+05.2 & 1.25 & 0.561 & 0.688 & 4261  \\
007.5+07.4 & 3    & 0.657 & 2.343 & 151   & 359.8+05.6 & 2.2  & 0.596 & 1.604 & 1403  \\
007.6+06.9 & 1.1  & 0.541 & 0.559 & 12955 & 359.8+06.9 & 1.7  & 0.578 & 1.122 & 1284  \\
007.8-03.7 & 2.2  & 0.596 & 1.604 & 862   & 359.9-04.5 & 1    & 0.528 & 0.472 & 16589 \\
\hline \hline
\end{tabular}
\label{T:hr}
\end{table*}

\end{document}